\documentclass[
  journal=pasa,
  manuscript=research-paper,
  year=2021,
  volume=37,
]{cup-journal}

\usepackage{microtype,siunitx,booktabs}
\usepackage{relsize}
\usepackage{amssymb}
\usepackage{amsmath}
\usepackage{hyperref}
\usepackage{upgreek}
\usepackage{dutchcal}
\usepackage{xcolor}
\definecolor{colorChange}{rgb}{0.7,0.0,0.0}

\newcommand{\ion}[2]{\ensuremath{\text{#1\hspace{0.15em}{\smaller #2}}}}
\newcommand{\degr}[0]{\ensuremath{^{\circ}}}
\newcommand{\code}[1]{\texttt{#1}}
\newcommand{\sun}[0]{\ensuremath{\odot}}

\newcommand{\nKin}[0]{124}        
\newcommand{\nKinUnique}[0]{109}  
\newcommand{\nKinOverlap}[0]{15}  

\newcommand{\ICRAR}[0]{International Centre for Radio Astronomy Research (ICRAR), The University of Western Australia, 35 Stirling Highway, Crawley WA 6009, Australia}
\newcommand{\ASTRO}[0]{ARC Centre of Excellence for All Sky Astrophysics in 3~Dimensions (ASTRO~3D)}
\newcommand{\CSIROPERTH}[0]{CSIRO Space and Astronomy, PO Box 1130, Bentley WA 6102, Australia}
\newcommand{\CSIROSYDNEY}[0]{CSIRO Space and Astronomy, PO Box~76, Epping NSW~1710, Australia}
\newcommand{\HERZBERG}[0]{NRC Herzberg Astronomy and Astrophysics Research Centre, 5071 W.~Saanich Rd., Victoria BC V9E~2E7, Canada}
\newcommand{\AUSSRC}[0]{Australian SKA Regional Centre (AusSRC)}
\newcommand{\USYDNEY}[0]{Sydney Institute for Astronomy, The University of Sydney, 44--70 Rosehill Street, Redfern NSW 2016, Australia}
\newcommand{\SWINBURNE}[0]{Centre for Astrophysics and Supercomputing, Swinburne University of Technology, PO Box~218, Hawthorn VIC~3122, Australia}

\title{WALLABY Pilot Survey: Public release of \ion{H}{I} data for almost 600~galaxies from phase~1 of ASKAP pilot observations}

\author{T.~Westmeier}
\affiliation{\ICRAR}
\alsoaffiliation{\ASTRO}
\email[T.~Westmeier]{tobias.westmeier@uwa.edu.au}

\author{N.~Deg}
\affiliation{Department of Physics, Engineering Physics, and Astronomy, Queen's University, Kingston ON K7L~3N6, Canada}

\author{K. Spekkens}
\affiliation{Department of Physics and Space Science, Royal Military College of Canada, P.O.\ Box 17000, Station Forces Kingston ON K7K~7B4, Canada}

\author{T.~N.~Reynolds}
\affiliation{\ICRAR}
\alsoaffiliation{\ASTRO}

\author{A.~X.~Shen}
\affiliation{\CSIROPERTH}
\alsoaffiliation{\AUSSRC}

\author{S.~Gaudet}
\affiliation{\HERZBERG}

\author{S.~Goliath}
\affiliation{\HERZBERG}

\author{M.~T.~Huynh}
\affiliation{\CSIROPERTH}
\alsoaffiliation{\ICRAR}

\author{P.~Venkataraman}
\affiliation{CIRADA, Dunlap Institute for Astronomy and Astrophysics, University of Toronto, Toronto ON M5S~3H4, Canada}

\author{X.~Lin}
\affiliation{Department of Astronomy, School of Physics, Peking University, 
Beijing 100871, China}

\author{T.~O'Beirne}
\affiliation{\ICRAR}
\alsoaffiliation{\ASTRO}

\author{B.~Catinella}
\affiliation{\ICRAR}
\alsoaffiliation{\ASTRO}

\author{L.~Cortese}
\affiliation{\ICRAR}
\alsoaffiliation{\ASTRO}

\author{H.~D\'{e}nes}
\affiliation{Netherlands Institute for Radio Astronomy (ASTRON), Oude Hoogeveensedijk~4, 7991 PD Dwingeloo, The Netherlands}

\author{A.~Elagali}
\affiliation{Telethon Kids Institute, Perth Children's Hospital, Perth, Australia}

\author{B.-Q.~For}
\affiliation{\ICRAR}
\alsoaffiliation{\ASTRO}

\author{G.~I.~G.~J\'{o}zsa}
\affiliation{Max-Planck-Institut f\"{u}r Radioastronomie, Radioobservatorium Effelsberg, Max-Planck-Stra\ss{}e~28, 53902~Bad M\"{u}nstereifel, Germany}
\alsoaffiliation{Department of Physics and Electronics, Rhodes University, PO Box~94, Makhanda, 6140, South Africa}

\author{C.~Howlett}
\affiliation{School of Mathematics and Physics, The University of Queensland, Brisbane QLD~4072, Australia}

\author{J.~M.~van der Hulst}
\affiliation{Kapteyn Astronomical Institute, University of Groningen, Landleven~12,
9747~AD~Groningen, The Netherlands}

\author{R.~J.~Jurek}
\affiliation{Unaffiliated}

\author{P.~Kamphuis}
\affiliation{Ruhr University Bochum, Faculty of Physics and Astronomy, Astronomical Institute, 44780~Bochum, Germany}

\author{V.~A.~Kilborn}
\affiliation{\SWINBURNE}
\alsoaffiliation{\ASTRO}

\author{D.~Kleiner}
\affiliation{INAF -- Osservatorio Astronomico di Cagliari, Via della Scienza~5, 09047~Selargius, CA, Italy}

\author{B.~S.~Koribalski}
\affiliation{\CSIROSYDNEY}
\alsoaffiliation{School of Science, Western Sydney University, Locked Bag~1797, Penrith NSW~2751, Australia}

\author{K.~Lee-Waddell}
\affiliation{\ICRAR}
\alsoaffiliation{\CSIROPERTH}

\author{C.~Murugeshan}
\affiliation{\CSIROPERTH}
\alsoaffiliation{\ASTRO}

\author{J.~Rhee}
\affiliation{\ICRAR}
\alsoaffiliation{\ASTRO}

\author{P.~Serra}
\affiliation{INAF -- Osservatorio Astronomico di Cagliari, Via della Scienza~5, 09047~Selargius, CA, Italy}

\author{L.~Shao}
\affiliation{National Astronomical Observatories, Chinese Academy of Sciences, 20A~Datun Road, Chaoyang District, Beijing~100101, China}

\author{L.~Staveley-Smith}
\affiliation{\ICRAR}
\alsoaffiliation{\ASTRO}

\author{J.~Wang}
\affiliation{Kavli Institute for Astronomy and Astrophysics, Peking University, Beijing~100871, China}

\author{O.~I.~Wong}
\affiliation{\CSIROPERTH}
\alsoaffiliation{\ICRAR}
\alsoaffiliation{\ASTRO}

\author{M.~A.~Zwaan}
\affiliation{European Southern Observatory, Karl-Schwarzschild-Stra\ss{}e~2, 85748~Garching, Germany}

\author{J.~R.~Allison}
\affiliation{First Light Fusion Ltd., Unit~9/10 Oxford Industrial Park, Mead Road, Yarnton, Kidlington OX5~1QU, UK}
\alsoaffiliation{\CSIROSYDNEY}

\author{C.~S.~Anderson}
\affiliation{Jansky Fellow of the National Radio Astronomy Observatory, P.~O.~48
Box~0, Socorro, NM~87801, USA}

\author{Lewis Ball}
\affiliation{\CSIROSYDNEY}
\alsoaffiliation{SKA Observatory, Jodrell Bank, Lower Withington, Macclesfield, Cheshire SK11~9FT, UK}

\author{D.~C.-J.~Bock}
\affiliation{\CSIROSYDNEY}

\author{D.~Brodrick}
\affiliation{Advanced Instrumentation Technology Centre, Research School of Astronomy and Astrophysics, Australian National University, Canberra ACT 2611, Australia}
\alsoaffiliation{\CSIROSYDNEY}

\author{J.~D.~Bunton}
\affiliation{\CSIROSYDNEY}

\author{F.~R.~Cooray}
\affiliation{\CSIROSYDNEY}

\author{N.~Gupta}
\affiliation{Inter-University Centre for Astronomy and Astrophysics, Post Bag~4, Ganeshkhind, Pune 411~007, India}
\alsoaffiliation{\CSIROSYDNEY}

\author{D.~B.~Hayman}
\affiliation{\CSIROSYDNEY}

\author{E.~K.~Mahony}
\affiliation{\CSIROSYDNEY}

\author{V.~A.~Moss}
\affiliation{\CSIROSYDNEY}
\alsoaffiliation{\USYDNEY}

\author{A.~Ng}
\affiliation{\CSIROSYDNEY}

\author{S.~E.~Pearce}
\affiliation{\CSIROSYDNEY}
\alsoaffiliation{SKA Observatory, 26~Dick Perry Avenue, Kensington WA~6151, Australia}

\author{W.~Raja}
\affiliation{\CSIROSYDNEY}

\author{D.~N.~Roxby}
\affiliation{\CSIROSYDNEY}

\author{M.~A.~Voronkov}
\affiliation{\CSIROSYDNEY}

\author{K.~A.~Warhurst}
\affiliation{CSIRO, 33~Onslow Street, Geraldton WA 6530, Australia}

\author{H.~M.~Courtois}
\affiliation{University of Lyon, UCB Lyon~1, CNRS/IN2P3, IP2I Lyon, France}

\author{K.~Said}
\affiliation{School of Mathematics and Physics, The University of Queensland, Brisbane QLD~4072, Australia}

\doi{to be added}

\received {dd Mmm YYYY}
\revised  {dd Mmm YYYY}
\accepted {dd Mmm YYYY}
\published{dd Mmm YYYY}

\keywords{radio lines: galaxies -- surveys -- catalogues} 

\begin{document}

\begin{abstract}
    We present WALLABY pilot data release~1, the first public release of \ion{H}{I} pilot survey data from the Wide-field ASKAP L-band Legacy All-sky Blind Survey (WALLABY) on the Australian Square Kilometre Array Pathfinder. Phase~1 of the WALLABY pilot survey targeted three $60~\mathrm{deg}^{2}$ regions on the sky in the direction of the Hydra and Norma galaxy clusters and the NGC~4636 galaxy group, covering the redshift range of $z \lesssim 0.08$. The source catalogue, images and spectra of nearly 600 extragalactic \ion{H}{I} detections and kinematic models for \nKinUnique{} spatially resolved galaxies are available. As the pilot survey targeted regions containing nearby group and cluster environments, the median redshift of the sample of $z \approx 0.014$ is relatively low compared to the full WALLABY survey. The median galaxy \ion{H}{I} mass is $2.3 \times 10^{9}~M_{\sun}$. The target noise level of $1.6~\mathrm{mJy}$ per $30''$ beam and $18.5~\mathrm{kHz}$ channel translates into a $5 \sigma$ \ion{H}{I} mass sensitivity for point sources of about $5.2 \times 10^{8} \, (D_{\rm L} / \mathrm{100~Mpc})^{2} \, M_{\sun}$ across 50~spectral channels ($\approx 200~\mathrm{km \, s}^{-1}$) and a $5 \sigma$ \ion{H}{I} column density sensitivity of about $8.6 \times 10^{19} \, (1 + z)^{4}~\mathrm{cm}^{-2}$ across 5~channels ($\approx 20~\mathrm{km \, s}^{-1}$) for emission filling the $30''$ beam. As expected for a pilot survey, several technical issues and artefacts are still affecting the data quality. Most notably, there are systematic flux errors of up to several 10\% caused by uncertainties about the exact size and shape of each of the primary beams as well as the presence of sidelobes due to the finite deconvolution threshold. In addition, artefacts such as residual continuum emission and bandpass ripples have affected some of the data. The pilot survey has been highly successful in uncovering such technical problems, most of which are expected to be addressed and rectified before the start of the full WALLABY survey.
\end{abstract}

\section{Introduction}

The Wide-field ASKAP L-band Legacy All-sky Blind Survey (WALLABY; \citealt{Koribalski2020}) is the all-sky \ion{H}{I} survey on the Australian Square Kilometre Array Pathfinder (ASKAP; \citealt{Hotan2021}). WALLABY was originally designed to detect the \ion{H}{I} emission of about half a million galaxies across three quarters of the sky south of a declination of $\delta \approx +30\degr$ and out to a redshift of $z \approx 0.26$. The lower-than-expected sensitivity of the ASKAP phased array feeds in combination with strong contamination of the frequency band below about $1300~\mathrm{MHz}$ by radio frequency interference from satellites mandated some adjustments to the sky and redshift coverage of the survey.

Following the recent formal review of ASKAP survey science projects, WALLABY has now been allocated 8832~hours of observing time over the first five years of full ASKAP operations. This will enable us to map approximately $1.4 \uppi~\mathrm{sr}$ of the sky at the original WALLABY sensitivity over the next five years and is expected to yield approximately 210,000 galaxies across a substantial fraction of the southern sky and out to a redshift of $z \approx 0.1$. Despite these adjustments, this number still corresponds to almost a ten-fold increase in the number of galaxies and a factor of 7 improvement in angular resolution compared to the state-of-the-art Arecibo Legacy Fast ALFA Survey (ALFALFA; \citealt{Haynes2018}) and is expected to create the largest legacy sample of \ion{H}{I}-detected galaxies in the local Universe until the start of regular science observations with the Square Kilometre Array (SKA; \citealt{Dewdney2009}).

WALLABY's science goals cover a wide range of astrophysical topics related to different aspects of galaxy formation and evolution \citep{Koribalski2020}. Most notably, WALLABY will enable the most thorough statistical analysis to date of the \ion{H}{I} content of galaxies in the local Universe as a function of environment \citep{Casasola2004,Doyle2005,Davies2019,Cortese2021,Saintonge2022}, including the measurement of \ion{H}{I} scaling relations (e.g., \citealt{Catinella2010,Catinella2018,Boselli2014,Denes2016,Casasola2020}), the \ion{H}{I} mass and velocity width functions (e.g., \citealt{Zwaan2005,Jones2016,Jones2020}) as well as the rotation curves (e.g., \citealt{deBlok2008,Kamphuis2015,Oh2018}) of several thousand galaxies expected to be sufficiently resolved by WALLABY's $30''$ angular resolution. In addition, WALLABY will be able to measure and refine several cosmological parameters and enable tests of General Relativity and the $\Lambda$CDM cosmological model (e.g., \citealt{Avila2021}; also see \citealt{Koribalski2020} and references therein).

Prior to full survey observations, we have undertaken 21-cm spectral line pilot surveys to check the performance of the ASKAP system and test the WALLABY post-processing and data analysis pipelines. Phase~1 of the pilot survey presented here targeted a total area of about $180~\mathrm{deg}^{2}$ (corresponding to six ASKAP pointings) on the sky split across three separate target fields in the direction of the Hydra and Norma galaxy clusters and the NGC~4636 galaxy group. The calibrated \ion{H}{I} data cubes of all fields produced by the ASKAP data reduction pipeline were made publicly available via the CSIRO ASKAP Science Data Archive (CASDA; \citealt{Huynh2020}) immediately after successful quality control. These raw image cubes are quite large ($\approx 0.5~\mathrm{TB}$ each) and, as the individual footprints are not combined and mosaicked together, they are difficult to handle and analyse.

The purpose of this first public WALLABY pilot data release (WALLABY PDR1) is to make available the value-added data products produced from the combined and mosaicked data cubes by the WALLABY team. These include source catalogues and source data products (data cubes, moment maps, spectra etc.) for almost 600 \ion{H}{I} sources detected by our source finding pipeline as well as kinematic models for over 100 spatially resolved galaxies. This paper focuses on the data processing, source finding and public release of the $\approx 600$ \ion{H}{I} sources, while a companion paper \citep{Deg2022} details the kinematic models applied to the spatially resolved detections as well as their public release.

In Section~\ref{sec_field_selection} we introduce the three target fields and provide an overview of their basic properties. Section~\ref{sec_observations} describes the observational setup and data reduction procedure. In Section~\ref{sec_data_quality} we evaluate the ASKAP data quality and discuss several technical issues that have affected the overall quality of the data. Section~\ref{sec_source_finding} outlines our blind \ion{H}{I} source finding and source characterisation approach followed by an overview of the basic parameters of the detected sources. In Section~\ref{sec_flux} we compare the ASKAP flux measurements with previous \ion{H}{I} surveys. Section~\ref{sec_kin} gives a brief overview of the kinematic models applied to suitable spatially-resolved sources. Instructions on how to retrieve the released data products are provided in Section~\ref{sec_data_access} followed by a brief summary in Section~\ref{sec_summary}.

\section{Field selection}
\label{sec_field_selection}

For phase~1 of WALLABY pilot survey observations we selected three fields in the direction of different galaxy group and cluster environments: the Hydra galaxy cluster, the Norma galaxy cluster and the NGC~4636 galaxy group. These targets were selected not purely for their scientific merit, but also for strategic reasons (detailed below) and because of their usefulness in testing our data reduction and analysis strategy.

At a distance of about 60~Mpc \citep{Jorgensen1996,Reynolds2022} the Hydra galaxy cluster (Abell~1060) is one of the nearest galaxy clusters. It has a virial mass of approximately $2 \times 10^{14} \, h_{70}^{-1} \, M_{\sun}$ \citep{Tamura2000}, and the cluster core is dominated by two massive elliptical galaxies, NGC~3309 and NGC~3311. Recent optical studies have found evidence for environmental effects such as star formation quenching \citep{LimaDias2021} and a reddening of galaxies towards the cluster centre \citep{LaMarca2022}, making it an interesting target for \ion{H}{I} observations. Parts of the core of the cluster were previously imaged in \ion{H}{I} emission with the VLA \citep{McMahon1993}.

The Norma cluster is located in the direction of the Great Attractor at a recession velocity of about $4900~\mathrm{km \, s}^{-1}$ \citep{Mutabazi2021}. With a dynamical mass of about $10^{15}~M_{\sun}$ \citep{Woudt2008} it is significantly more massive than the Hydra cluster. The cluster is home to the famous “jellyfish galaxy”, ESO~137$-$001, which features an up to 70~kpc long tail of X-ray, H$\upalpha$ and CO emission \citep{Sun2006,Sun2007,Jachym2019} caused by ram-pressure stripping. Its low Galactic latitude makes the Norma cluster a challenging target for optical studies, and \ion{H}{I} observations can therefore make an important contribution to studying the galaxies in and around the cluster, although few targeted \ion{H}{I} observations have been attempted so far \citep{KraanKorteweg1997}.

The nearby NGC~4636 group is located close to the celestial equator in the southern outskirts of the Virgo cluster. It was chosen to facilitate joint \ion{H}{I} observations with the Five-hundred-meter Aperture Spherical Radio Telescope (FAST; \citealt{Nan2011}) to enable better quantification of the diffuse \ion{H}{I} emission \citep{Zuo2022}. The group is dominated by the giant elliptical galaxy NGC~4636 at a distance of about 15~Mpc \citep{Tully2013} which is known for its extended and complex X-ray halo \citep{Jones2002,Baldi2009}. NGC~4636 has also been detected in \ion{H}{I} emission, implying an \ion{H}{I} mass of approximately $8.2 \times 10^{8}~M_{\sun}$ \citep{Knapp1978}.

\section{Observations and data reduction}
\label{sec_observations}

\begin{figure}[t]
    \centering
    \includegraphics[width=\linewidth]{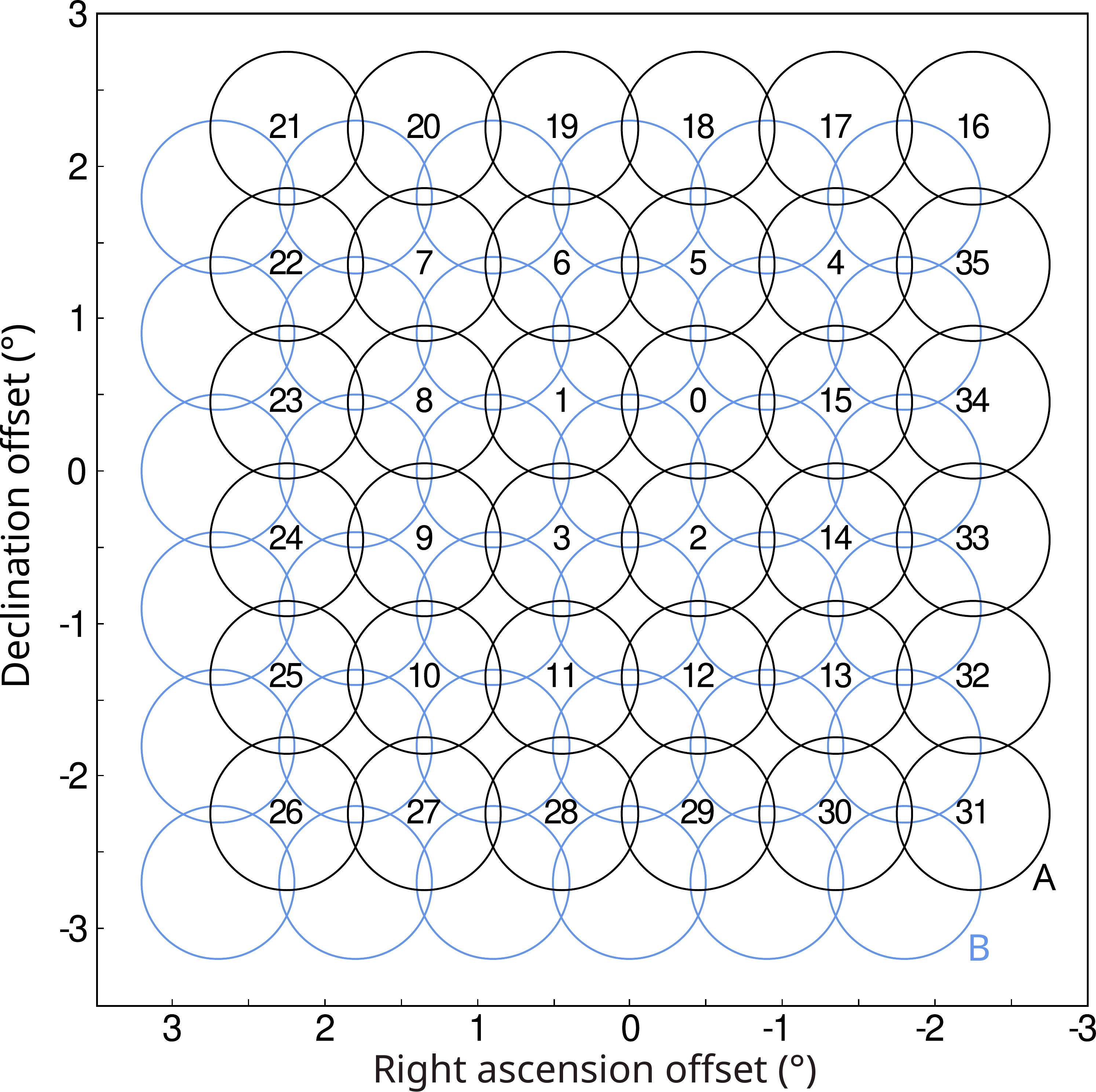}
    \caption{Each WALLABY tile on the sky is composed of two interleaved footprints referred to as A and B (shown in black and blue, respectively). In the `\texttt{square\_6x6}' configuration, each footprint is made up of 36~instantaneous primary beams (labelled with numbers for footprint~A here) separated by $0.9\degr$. The beam size is $\lambda / D \approx 1\degr$ for an antenna of diameter $D = 12~\mathrm{m}$ at a wavelength of $\lambda = 21.1~\mathrm{cm}$.}
    \label{fig_askap_footprint}
\end{figure}

Each field covers about $60~\mathrm{deg}^{2}$ of sky in the form of two adjacent $30~\mathrm{deg}^{2}$ ASKAP tiles. Each tile is composed of two interleaved 36-beam ASKAP footprints to achieve a more uniform sensitivity across the field. Each footprint was observed in the `\texttt{square\_6x6}' configuration which arranges the 36~primary compound beams of the Phased Array Feed in a square pattern of $6 \times 6$~beams with an angular separation of $0.9\degr$ between two adjacent primary beams (Fig.~\ref{fig_askap_footprint}). The on-source integration time was approximately $8~\mathrm{h}$ per footprint and thus $16~\mathrm{h}$ per tile. The total on-source observing time for the three fields of the WALLABY phase~1 pilot survey therefore amounts to approximately $6 \times 16~\mathrm{h}$ which corresponds to just under 100~hours. A summary of the observations is presented in Table~\ref{tab_observations}. In addition, the standard calibrator PKS~1934$-$638 was typically observed once a day for a total of about 2.5~hours to obtain data for flux and bandpass calibration of each of the 36~ASKAP primary beams.

\begin{table*}[t]
    \centering
    \caption{Summary of WALLABY phase~1 pilot survey observations. The columns denote the name of the field, the tile/footprint number, the date of the observation, the pointing centre ($\alpha_{\rm J2000}, \delta_{\rm J2000}$) in equatorial coordinates, the rotation angle ($\phi$) of the footprint on the sky, the on-source integration time ($t_{\rm int}$), the official ASKAP scheduling block identifier (SBID) which can be used to locate the data in CASDA, and any comments on the quality of the data.}
    \label{tab_observations}
    \begin{tabular}{lllrrrrrl}
        \hline
        Field & Tile & Date & $\alpha_{\rm J2000}$ & $\delta_{\rm J2000}$ & $\phi$  & $t_{\rm int}$ & SBID & Comments \\
              &      &      & (\degr)              & (\degr)              & (\degr) & (h)           &     &          \\
        \hline
		Hydra    & 1A & 2019 Nov 20 & $153.9494$ & $-27.3744$ & $-0.004$ & $8.0$ & $10609$ & Beam 26 discarded due to on-dish calibrator ripple. \\
		         & 1B & 2019 Oct 25 & $154.4582$ & $-27.8234$ & $-0.240$ & $8.0$ & $10269$ & Antennas 16 and 31 flagged. \\
		         & 2A & 2019 Nov 21 & $159.8510$ & $-27.3744$ & $-0.004$ & $8.0$ & $10612$ & Beams 21 and 26 discarded due to on-dish calibrator ripple. \\
		         & 2B & 2019 Nov 23 & $160.3598$ & $-27.8234$ & $-0.240$ & $8.0$ & $10626$ & \\
		\hline
		NGC~4636 & 1A & 2019 Dec 12 & $189.5097$ &  $-0.4500$ & $ 0.000$ & $8.7$ & $10809$ & Continuum residuals from 3C~273. \\
		         & 1B & 2019 Dec 13 & $189.9597$ &  $-0.9000$ & $-0.010$ & $8.7$ & $10812$ & Continuum residuals from 3C~273. \\
		         & 2A & 2019 Dec 5  & $189.5114$ &   $4.9496$ & $ 0.001$ & $8.7$ & $10736$ & Continuum residuals from 3C~273; beams 12 and 31 flagged. \\
		         & 2B & 2019 Dec 7  & $189.9627$ &   $4.4995$ & $ 0.040$ & $8.7$ & $10756$ & Discarded due to poor data quality. \\
		\hline
		Norma    & 1A & 2020 Feb 14 & $244.1289$ & $-59.4616$ & $-0.013$ & $8.1$ & $11816$ & Continuum residuals from  ESO~137$-$G006. \\
		         & 1B & 2020 Feb 15 & $245.0264$ & $-59.9086$ & $-0.770$ & $8.1$ & $11832$ & Continuum residuals from  ESO~137$-$G006. \\
		         & 2A & 2020 Mar 13 & $253.8586$ & $-59.4616$ & $-0.013$ & $8.1$ & $12193$ & \\
		         & 2B & 2020 Mar 14 & $254.7561$ & $-59.9086$ & $-0.770$ & $8.1$ & $12209$ & \\
        \hline
    \end{tabular}
\end{table*}

Observations were carried out in band~2 of ASKAP which covers the frequency range of 840 to 1440~MHz (see Table~3 in \citealt{Hotan2021}). The correlator was set up to supply 15,552~spectral channels across an instantaneous bandwidth of 288~MHz in the frequency range of 1151.5 to 1439.5~MHz. Unfortunately, the lower half of the band below $1300~\mathrm{MHz}$ is strongly affected by radio frequency interference (RFI) caused by global navigation satellite systems (GNSS). While there may be narrow windows that are less severely affected by RFI, it is not generally possible to obtain scientifically useful data from the lower half of the band. We therefore chose to only calibrate and image data from the upper 144~MHz of the band in the range of 1295.5 to 1439.5~MHz, thus limiting the redshift range across which the \ion{H}{I} emission line is in principle detectable to $z < 0.096$.

The data were reduced and calibrated using the ASKAPsoft data reduction pipeline \citep{Guzman2019,Whiting2020}. Each of the 36 instantaneous primary beams of an observation was reduced independently in ASKAPsoft. The data reduction procedure involved automatic flagging of bad data, bandpass calibration derived from observations of the standard calibrator, PKS~1934$-$638, and gain calibration via self-calibration of the radio continuum sources detected within the primary beam. The continuum emission was then subtracted from the visibility data using a sky model derived from the calibrated and deconvolved continuum image.

The calibrated and continuum-subtracted visibility data for all baselines up to a maximum length of $2~\mathrm{km}$ were imaged using robust weighting \citep{Briggs1995} with a robustness parameter of 0.5 and additional Gaussian tapering to obtain a synthesised beam size of close to $30''$ (using a spatial pixel size of $6''$). This is achieved through pre-conditioning (i.e.\ image-plane weighting) in ASKAPsoft rather than traditional weighting in visibility space. A second round of continuum subtraction was then carried out in the image domain to try to fit and remove any residual continuum emission.

The imager in ASKAPsoft also carries out basic Doppler correction from the topocentric to the barycentric reference frame. This is currently done by calculating the correction for each phase centre at the beginning of an observation and then applying that constant correction across the entire duration of the observation. For an 8-hour integration the resulting error within each primary beam is negligible and significantly smaller than the $18.5~\mathrm{kHz}$ channel width. Spectra are shifted using nearest-neighbour interpolation, which could introduce systematic frequency shifts even larger than the Doppler errors, although such shifts are still insignificant compared to the statistical uncertainties of WALLABY's frequency centroid measurements.

Next, the spectral image cubes for each primary beam were deconvolved using an implementation of the multi-scale CLEAN algorithm \citep{Cornwell2008,Rau2011} with spatial scales of 0, 3, 10 and 30~pixels. We applied major-cycle and minor-cycle CLEAN thresholds of $0.5~\mathrm{mJy}$ and $3.5~\mathrm{mJy}$, respectively, with an additional deep CLEAN threshold equal to the major-cycle threshold. After deconvolution, all 36~primary beams were mosaicked together into a single footprint (under the simplified assumption that there is no statistical correlation between adjacent beams), and the two 8-hour footprints for each targeted region on the sky were finally mosaicked into a single 72-beam tile of full sensitivity ($\approx 1.6~\mathrm{mJy}$ per spectral channel and beam) and adequate dithering for optimal noise uniformity across the $30~\mathrm{deg}^{2}$ region covered by each tile.

More detailed information on the WALLABY data reduction procedure is available in \citet{LeeWaddell2019} and \citet{Kleiner2019}.

\section{Data quality and limitations}
\label{sec_data_quality}

One of the main motivations for commissioning observations with new telescopes is the need to identify technical issues that could adversely affect the scientific quality of the data. The WALLABY pilot survey has been successful in uncovering several issues with both the hardware and the data reduction software pipeline of ASKAP. Several of these issues have since been rectified, which will result in a significant improvement in data quality for future WALLABY observations.

At just under 600, the total number of galaxies detected during phase~1 of the pilot survey across six $30~\mathrm{deg}^{2}$ tiles is significantly lower than the average of 490~galaxies per tile predicted by the WALLABY reference simulation presented in \citet{Koribalski2020}. However, this discrepancy is readily explained by a number of contributing factors, some of which are field-specific:
\begin{itemize}
    \item The reference simulation assumes a contiguous sky coverage with homogeneous sensitivity, whereas the noise in our pilot data significantly increases towards the edge of each field due to the lack of adjacent sky tiles (Section~\ref{sec_noise_and_sensitivity}).
    \item The reference simulation assumes a redshift cutoff of $z < 0.26$, whereas the pilot data only covered half of the bandwidth out to $z \approx 0.1$.
    \item The entire NGC~4636 field is severely affected by continuum residuals which prevented us from carrying out a blind source finding run and significantly reduced the number of \ion{H}{I} detections that could be extracted from the two tiles of that field. In addition, one of the two footprints making up the northern tile had to be discarded due to data quality issues, further reducing our sensitivity.
    \item Likewise, the Norma field was severely affected by continuum emission, and only one of its two tiles contained useful data.
    \item Low-level continuum emission (Section~\ref{sec_continuum}), bandpass ripples caused by ASKAP’s on-dish calibrators (Section~\ref{sec_odc}), correlator dropouts (Section~\ref{sec_correlator}), corrupt beams (Section \ref{sec_deconvolution}) and other artefacts have led to both a local reduction in sensitivity and difficulties in pushing the source finding pipeline as deep as originally intended, thus further reducing the number of \ion{H}{I} detections.
    \item Lastly, the pilot fields were selected to contain foreground clusters and groups at $z \lesssim 0.02$, but turned out to be under-dense at intermediate redshifts of $z \approx 0.04$ where -- due to the larger cosmological volume and lower system temperature -- the detection rate of the full WALLABY survey is expected to peak.
\end{itemize}
Several of these problems are discussed in more detail throughout the remainder of this section. Overall, the pilot survey has been highly successful in uncovering such data quality issues, thus giving us the opportunity to rectify most of them before the start of full WALLABY survey observations. We are therefore confident that the source detection rate of the full survey will be more consistent with the predictions presented in \citet{Koribalski2020}.

\subsection{Noise level and sensitivity}
\label{sec_noise_and_sensitivity}

Noise maps of all three fields are presented in Fig.~\ref{fig_noise_hydra}, \ref{fig_noise_n4636} and \ref{fig_noise_norma} in \ref{sec_noise_maps}. In general, we are able to achieve the nominal WALLABY noise level of $1.6~\mathrm{mJy}$ per $30''$ beam and $18.5~\mathrm{kHz}$ frequency channel across the inner regions of all tiles (typically the inner $4\degr \times 4\degr$ of a single tile) unless individual primary beams or entire footprints are missing. The noise does vary slightly across the sky depending on data quality and flagging fraction. While there is a strong increase in the noise level towards the edges of each field, this issue will not affect the full WALLABY survey which will have a contiguous sky coverage with adequate overlap between tiles.

\begin{figure}[t!]
    \centering
    \includegraphics[width=\linewidth]{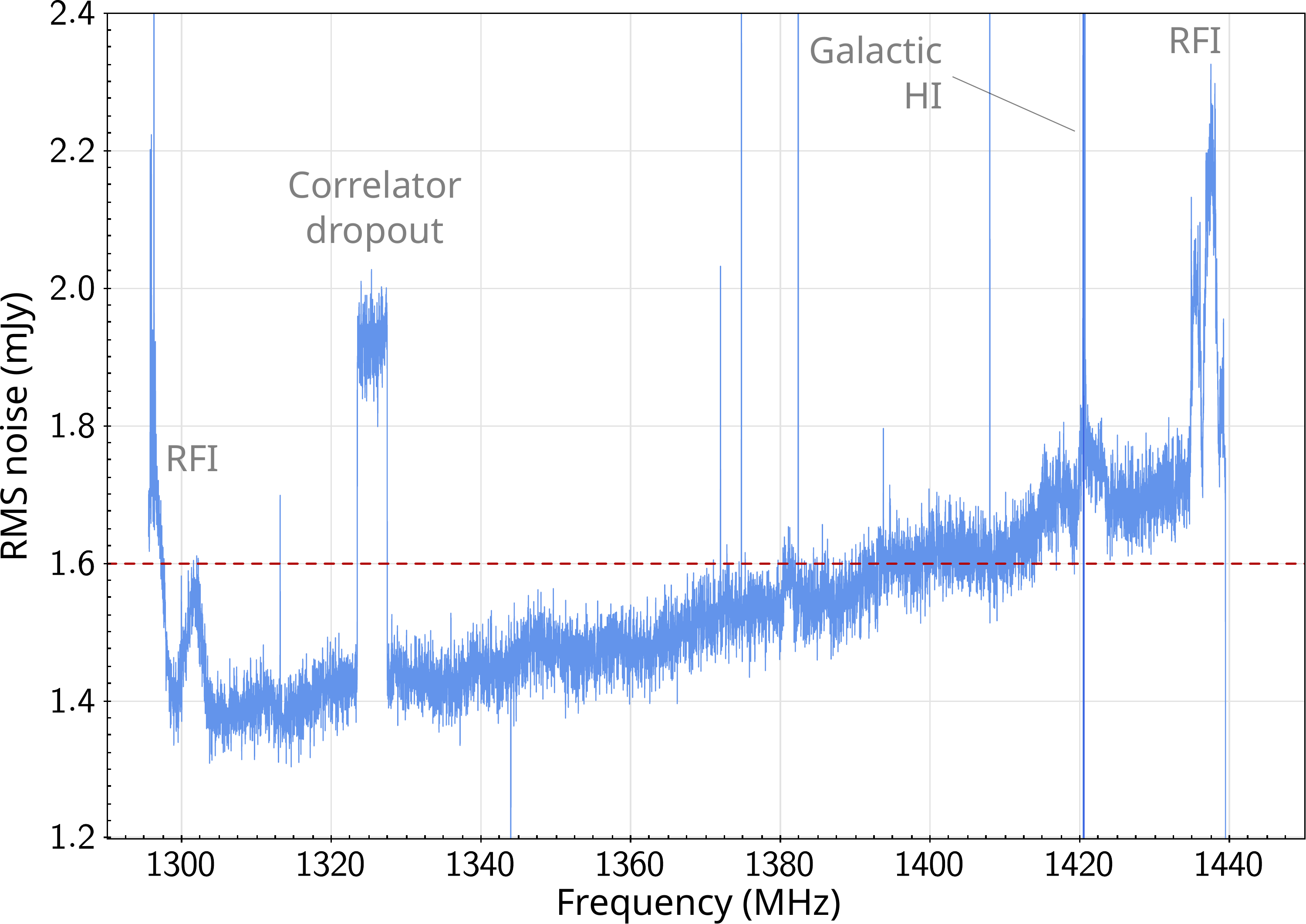}
    \caption{RMS noise level as a function of frequency in the eastern tile of the Norma field near $\alpha = 17^{\rm h} 03^{\rm m} 02^{\rm s}$ and $\delta = -59\degr 39' 05''$. The target WALLABY noise level of $1.6~\mathrm{mJy}$ is indicated by the dashed, red line.}
    \label{fig_noise}
\end{figure}

The noise level does vary as a function of frequency as well, as is shown in Fig.~\ref{fig_noise} for a location in the eastern tile of the Norma field. In particular, the noise is marginally higher ($\approx 1.7~\mathrm{mJy}$) at the low-redshift end of the band, but drops to $\approx 1.4~\mathrm{mJy}$ at the high-redshift end as a result of decreasing system temperature and increasing primary beam overlap when mosaicking multiple footprints. A similar behaviour is seen in all fields, although the noise level is generally much higher across the northern tile of the NGC~4636 field due to one footprint missing (see Fig.~\ref{fig_noise_n4636} in \ref{sec_noise_maps}).

The target noise level of WALLABY of $1.6~\mathrm{mJy}$ per beam and spectral channel translates into a $5 \sigma$ \ion{H}{I} mass detection limit of about $5.2 \times 10^{8} \, (D_{\rm L} / 100~\mathrm{Mpc})^{2} \, M_{\sun}$ over 50 spectral channels ($\approx 200~\mathrm{km \, s}^{-1}$), assuming a point source. The corresponding $5 \sigma$ \ion{H}{I} column density sensitivity is $8.6 \times 10^{19} \, (1 + z)^{4} \, \mathrm{cm}^{-2}$ over 5~channels ($\approx 20~\mathrm{km \, s}^{-1}$) for emission filling the $30''$ beam.

\subsection{Image deconvolution}
\label{sec_deconvolution}

In the current WALLABY data reduction pipeline, image deconvolution is carried out on each individual primary beam and prior to mosaicking of all 36~beams of a footprint and combination of two footprints into a single tile. As a result, faint sidelobes which are below the noise level in an individual primary beam may resurface after mosaicking due to the lower noise floor in the mosaicked tile. While this is not generally a problem for WALLABY pilot survey data due to the low intrinsic sidelobe levels of ASKAP's synthesised beam and the faintness of the majority of the sources detected by the survey, sidelobes were picked up by the source finding pipeline near individual bright and extended objects, most notably around the bright galaxy NGC~3137 (WALLABY J100903$-$290239) in the Hydra field.

Another problem affecting parts of the data is that the deconvolution algorithm appears to have failed to converge for a limited number of spectral channels and primary beams. This divergence occurs in channels where most of the data have been flagged, e.g.\ due to excessive RFI, and results in certain primary beams having arbitrary positive and negative flux density values in a small number of spectral channels. Due to their visual appearance these have been dubbed `zebra beams'. They pose a significant challenge to source finding, and our current approach is to automatically detect and flag entire spectral channels containing `zebra beams' at the start of the source finding pipeline. As a result, some galaxies may be missing flux if flagged channels occurred across their spectral profile width.

Another intrinsic issue related to deconvolution in general is the fact that only bright emission above the noise level can be deconvolved, while faint emission sitting below the noise level at the original spatial and spectral resolution of the data will remain convolved with the dirty beam. As a result, faint and bright sources in the same deconvolved image could be characterised by rather different beams (or possibly even a combination of the clean and dirty beam) which poses a challenge for the accurate recovery of the integrated flux of a source, in particular as the dirty beam will generally vary with frequency. This problem is not specific to WALLABY, but affects any blind interferometric survey. We discuss the resulting consequences in more detail in Section~\ref{sec_flux}.

This specific problem was first noted in the Hydra internal team release~1 (Hydra TR1) data which has a restoring beam size of $30''$, but a dirty beam size that is somewhat smaller and slightly elliptical, ranging from about $28.5'' \times 26''$ at $1295.5~\mathrm{MHz}$ to about $30'' \times 27.5''$ at $1439.5~\mathrm{MHz}$. As a result, the flux measurements for fainter, not fully deconvolved sources in Hydra TR1, which are based on the assumption of a $30''$ circular beam, are too low by as much as 18\% at the low-frequency end of the band.

\begin{table*}[t]
    \centering
    \caption{Important SoFiA parameter settings in the S+C finder (\code{scfind}), linker and reliability modules for the different phase~1 pilot survey fields. Hydra TR1 and TR2 are two separate source finding runs on the Hydra field (see Section~\ref{sec_hydra_sf}). All spatial (XY) and spectral (Z) parameters are given in pixels. The last row lists the resulting number of sources detected in each field.}
    \label{tab_sofiapar}
    \begin{tabular}{lrrrr}
        \hline
	    Parameter & Hydra TR1 & Hydra TR2 & NGC~4636 TR1 & Norma TR1 \\
	    \hline
        \code{scfind.kernelsXY} & 0, 5, 10 & 0, 5, 10 & 0, 5, 10 & 0, 5, 10 \\
        \code{scfind.kernelsZ} & 0, 3, 7, 15 & 0, 3, 7, 15 & 0, 3, 7, 15, 31 & 0, 3, 7, 15, 31 \\
        \code{scfind.threshold} & 3.5 & 3.5 & 4 & 3.5 \\
        \code{linker.radiusXY} & 2 & 2 & 2 & 2 \\
        \code{linker.radiusZ} & 2 & 2 & 2 & 2 \\
        \code{linker.minSizeXY} & 8 & 8 & 3 & 5 \\
        \code{linker.minSizeZ} & 5 & 5 & 3 & 8 \\
        \code{reliability.threshold} & 0.6 & 0.8 & 0.9 & 0.8 \\
        \code{reliability.scaleKernel} & 0.3 & 0.4 & 0.5 & 0.35 \\
        \code{reliability.minSNR} & 2.8$^{\dagger}$ & 2.8$^{\dagger}$ & 5 & 2.8$^{\dagger}$ \\
        \hline
        Detections & 148 & 272 & 147 & 144 \\
        \hline
        \multicolumn{5}{l}{$^{\dagger}\,$The old SoFiA parameter \code{reliability.fmin} = 15 was used, corresponding to \code{minSNR} $\approx$ 2.8.}
    \end{tabular}
\end{table*}

\subsection{Residual continuum emission}
\label{sec_continuum}

All phase~1 pilot survey data cubes still contain residual radio continuum emission as a result of imperfect continuum subtraction by the ASKAP data reduction pipeline. These residuals manifest themselves as a periodic bandpass ripple feature in the direction of bright radio continuum sources and their sidelobes. The severity of the continuum artefacts is a function of angular distance from the position of the continuum source and also depends on whether the continuum source is located inside the field (in which case it will have been imaged and subtracted by the data reduction pipeline) or not. In addition, due to the declination-dependence of ASKAP's $UV$ coverage, the amplitude of continuum residuals is expected to be generally higher near the celestial equator due to elevated beam sidelobe levels.

In WALLABY pilot data, continuum residuals become detectable when the flux density of the original continuum source exceeds $\approx 100~\mathrm{mJy}$. For brighter sources of $\gtrsim 1~\mathrm{Jy}$ the amplitude of the ripple becomes more significant, and faint sidelobes around the position of the continuum source may show up as well. Very bright sources of $\gtrsim 10~\mathrm{Jy}$ have catastrophic effects when the current pipeline is employed, as they create sidelobes and other artefacts across a significant portion of the sky area covered by the affected tile.

The Hydra field is the least affected one, with the brightest source, PKS~1006$-$286, having a flux density of about 1.5~Jy at 1.4~GHz \citep{Condon1998}. However, the other two fields both contain a very bright continuum source that affects a significant fraction of the data. In the case of the Norma field, the extremely bright and extended radio galaxy ESO~137$-$G006 \citep{Ramatsoku2020}, also known as PKS~B1610$-$607, is located just south of the centre of the western tile. At an integrated flux density of $\approx 43~\mathrm{Jy}$, it is one of the central galaxies of the Norma cluster itself and creates significant sidelobes and artefacts across almost the entire western tile. Another extended radio galaxy, ESO~137$-$G007, lies just $15'$ to the north of ESO~137$-$G006.

Likewise, the NGC~4636 field contains the bright quasar 3C~273 which is located in the north-western corner of the southern tile and the south-western corner of the northern tile. At a flux density of $\approx 55~\mathrm{Jy}$ \citep{Condon1998}, it creates severe artefacts across the entire $60~\mathrm{deg}^{2}$ field. The strong continuum artefacts in the Norma and NGC~4636 fields have limited our ability to extract an \ion{H}{I} source catalogue for those fields. In the case of Norma we were only able to run our source finding pipeline on the eastern tile which is unaffected by continuum residuals, while for the NGC~4636 field we resorted to searching for \ion{H}{I} emission only in small regions around the position and redshift of known galaxies. The latter was not possible for the western tile of Norma due to the low Galactic latitude of that field and the resulting lack of optical redshift measurements.

\subsection{On-dish calibrator}
\label{sec_odc}

Another artefact discovered in the data during source finding is a faint spectral ripple feature that can be seen in several of the ASKAP primary beams and is most prominent in some of the corner beams of the $6 \times 6$ beam footprint. Tests later established that this ripple was caused by the telescope’s on-dish calibrators (ODC) which were left switched on during the observations. Tests during phase~2 of ASKAP pilot observations have confirmed that the ripple disappears when the ODCs remain powered off during science observations.

In order to minimise the impact of the ODC ripple on our automated source finding, we identified the most severely affected primary beams in the Hydra field and removed them from the final mosaic. This results in the loss of sensitivity in small areas of the field around the three corner beams that had to be removed, but at the benefit of being able to obtain a much cleaner and overall more complete \ion{H}{I} source catalogue. We did not remove any primary beams from the Norma field, but instead developed and applied a ripple removal filter prior to source finding which helped in reducing the impact of the ODC. The presence of the ripple was not critical in the NGC~4636 field due to our local source finding approach in the direction of known optical sources, as discussed in Section~\ref{sec_ngc4636}.

\begin{figure*}[t!]
    \centering
    \includegraphics[width=\linewidth]{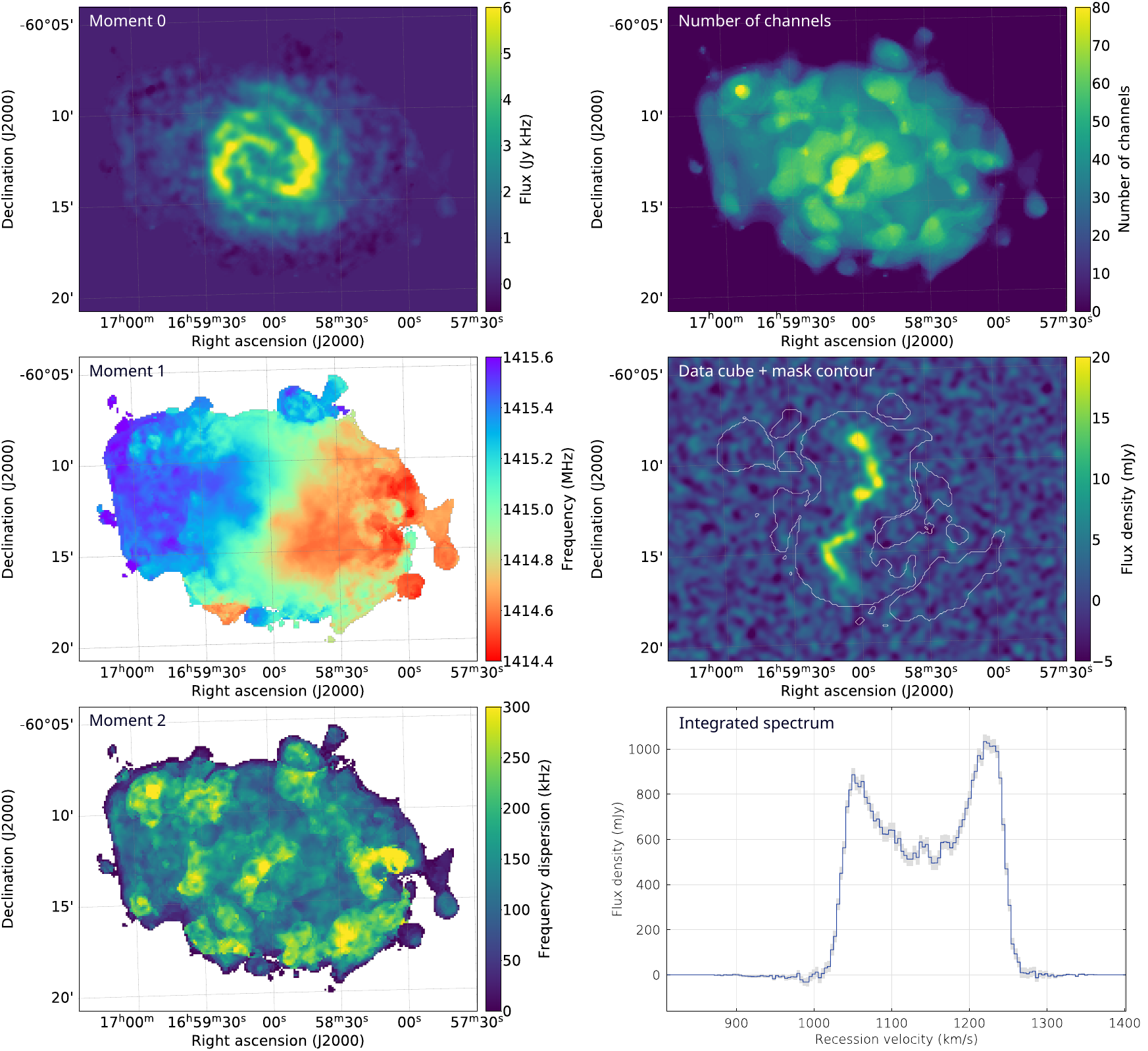}
    \caption{Example SoFiA output products for a bright, nearby galaxy (WALLABY J165901$-$601241 / ESO~138$-$G010 in the Norma field at $\mathrm{c}z \approx 1140~\mathrm{km \, s}^{-1}$), showing maps of the $0^{\rm th}$, $1^{\rm st}$ and $2^{\rm nd}$ spectral moment (left column), a map of the number of spectral channels per pixel (upper-right), a single channel map of the data cube with associated mask (centre-right) and the integrated spectrum with statistical uncertainties shaded in grey (bottom-right).}
    \label{fig_example_output}
\end{figure*}

\begin{figure*}[t!]
    \centering
    \includegraphics[width=0.87\linewidth]{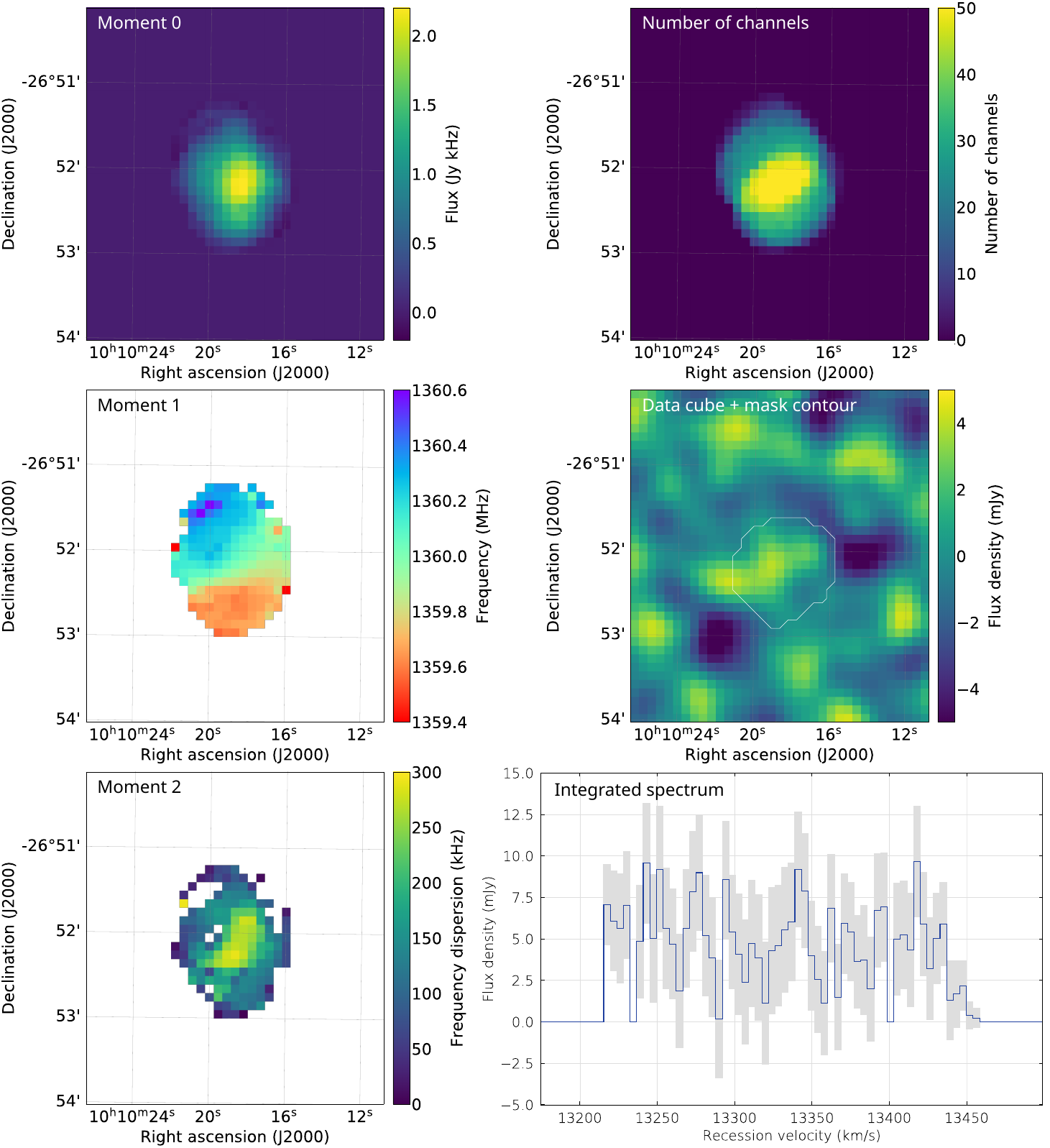}
    \caption{Same as Fig.~\ref{fig_example_output}, but for a faint, distant galaxy (WALLABY J101018$-$265209 / LEDA~762864 in the Hydra field at $\mathrm{c}z \approx 13300~\mathrm{km \, s}^{-1}$).}
    \label{fig_example_output_2}
\end{figure*}

\subsection{Correlator drop-outs}
\label{sec_correlator}

Several of the phase~1 footprints are affected by individual correlator drop-outs whereby one or two 4~MHz correlator modules appear to have failed during the observation. This results in missing data across certain 4~MHz frequency intervals in a footprint and thus an increased noise level (by a factor of $\sqrt{2}$) in the final mosaic across the affected intervals due to only one of the two footprints containing valid data. While this does not have a severe impact on the scientific quality of the data, it does reduce our sensitivity to \ion{H}{I} emission in small redshift intervals affected by the drop-outs (see Fig.~\ref{fig_noise} for an example at $\approx 1325~\mathrm{MHz}$).

\subsection{Primary beam shape}

During phase~1 of WALLABY pilot observations, information on the exact shape of the 36~primary beams provided by the ASKAP phased array feed was unavailable. In the mosaicking process, all primary beams were therefore assumed to be of Gaussian shape with a frequency-dependent FWHM of roughly $1\degr$ at $1.4~\mathrm{GHz}$. This will naturally introduce a certain level of flux uncertainty, in particular in the outer beams which are expected to deviate more significantly from a Gaussian shape \citep{Serra2015a,Hotan2021}.

Tests with improved primary beam models derived from holographic measurements \citep{Hotan2016} have proven successful, and holography beam models instead of Gaussian beams will be used in all future WALLABY observations. A comparison of \ion{H}{I} fluxes derived from Gaussian and holography beam mosaics has shown that the holography fluxes are on average slightly smaller than the Gaussian ones, as the holography beams are somewhat more extended than the $\approx 1 \degr$ Gaussian beams assumed during phase~1. For an individual footprint the flux discrepancy can be as large as 15\% depending on the angular separation of the source from the nearest beam centres, although the effect is much smaller (less than 5\%) for the final tiles of two interleaved footprints. Hence, we do not expect the fluxes in WALLABY phase~1 pilot survey data to be significantly affected by our choice of primary beam model.

\section{Source finding and parameterisation}
\label{sec_source_finding}

Source finding on all fields was carried out on the in-house Hyades computing cluster at the International Centre for Radio Astronomy Research (ICRAR) in Perth using version~2 of the Source Finding Application (SoFiA; \citealt{Serra2015,Westmeier2021}).\footnote{\url{https://github.com/SoFiA-Admin/SoFiA-2}} In order to speed up the source finding process, the SoFiA-X environment\footnote{\url{https://github.com/AusSRC/SoFiAX}} was used to process smaller subsections of the full data cube in parallel on multiple Hyades nodes.

\subsection{Preconditioning}

At the start of the source finding process, each data cube was multiplied by the square root of the weights cube supplied by the ASKAP data reduction pipeline to normalise noise levels across the entire cube. Noise variations would typically result from effects such as primary beam attenuation, system temperature variation, variable data flagging fraction or missing data in specific primary beams and/or frequency ranges. In addition to applying the weights cube, we also used SoFiA's built-in noise normalisation module to measure and divide by the local noise level in a running window of size $51 \times 51$ spatial pixels and 51~spectral channels. This ensures that the noise is constant throughout the entire data cube to allow a constant source detection threshold to be applied later on.

Next, SoFiA's auto-flagger was used to automatically flag spatial pixels or spectral channels for which $|\sigma_{i} - \mu| > 5 \mathrm{k} \tilde{\mu}$, where $\sigma_{i}$ is the normalised RMS noise level (measured robustly using the median absolute deviation) in pixel or channel~$i$, $\mu$ is the median RMS noise level of all pixels or channels, $\tilde{\mu}$ is the median absolute deviation from $\mu$, and $\mathrm{k} = 1 / \Phi^{-1}(3/4) \approx 1.4826$ is a constant to convert between median absolute deviation and standard deviation under the assumption that the underlying values obey a normal distribution ($\Phi^{-1}$ is the inverse of the quantile function). With this setup, the auto-flagger removes severe artefacts in the data such as RFI, residual continuum emission and so-called `zebra beams' caused by the deconvolution algorithm failing to converge. The \ion{H}{I} emission from galaxies is not flagged, as they are spatially and spectrally compact.

In addition, we specifically flagged a radius of 5~pixels around the positions of all radio continuum sources brighter than about 100 to 150~mJy to remove faint bandpass ripples associated with weaker continuum sources that the auto-flagger would not have removed. Continuum source positions were either extracted from the NVSS catalogue \citep{Condon1998} where possible, or otherwise from the ASKAP radio continuum image of the field itself.

In the case of the eastern tile of the Norma field we additionally ran SoFiA's ripple removal filter to reduce the ODC-generated bandpass ripple that was noticeable in some of the edge and corner beams of the mosaic. The filter works by subtracting the median of the data from a running window, the size of which was set to $91 \times 91$ spatial pixels ($9'$) and 41~spectral channels ($760~\mathrm{kHz}$) to ensure that the filter was large enough to not be affacted by \ion{H}{I} emission from galaxies. We did not resort to the ripple filter for the Hydra field, but instead opted to remove the three primary beams that were most strongly affected by the ODC ripple, which is insignificant compared to the total number of 144 primary beams making up the full $60~\mathrm{deg}^{2}$ field. Likewise, ripple removal was not carried out on the NGC~4636 field either, as that field is affected by continuum artefacts which are far more severe than the ODC ripple.

\subsection{Source finding}

We used SoFiA's default source finding algorithm, the smooth-and-clip (S+C) finder, which smooths the data on multiple spatial and spectral scales chosen by the user, measures the noise level in each smoothing iteration and then applies a user-defined flux threshold relative to the noise to retain all pixels with a statistically significant flux density. For reasons of bias reduction, SoFiA also retains negative pixels with an absolute flux density greater than the threshold. The basic S+C finder settings used for each field, including the spatial and spectral smoothing scales and flux detection threshold, are given in Table~\ref{tab_sofiapar}.

The next step was to merge all of the detected pixels into coherent sources. This is done by SoFiA's linker module which merges pixels that are within a user-specified merging length of each other. The basic linker settings used for each field are also listed in Table~\ref{tab_sofiapar}.

The final step in SoFiA was the removal of false positives using the reliability module. Reliability filtering works by comparing the density of detections with net positive and negative flux in a user-defined parameter space and discarding all positive detections that have a reliability of less than a user-specified minimum threshold. The method is described in detail in \citet{Serra2012}. The basic settings used for the different fields are again given in Table~\ref{tab_sofiapar}.

After source finding, linking and reliability filtering, SoFiA writes out the resulting source catalogue and other relevant data products such as integrated source spectra and moment maps. Each detection was then visually assessed by inspecting the $0^{\rm th}$ and $1^{\rm st}$ moment, the integrated spectrum and an optical image from the Digitized Sky Survey (DSS) in an attempt to identify and discard any false positives caused by noise peaks or artefacts. 711~\ion{H}{I} sources remain in the catalogue after the manual removal of false positives and are included in this data release. The last row in Table~\ref{tab_sofiapar} lists the number of sources detected in each field.

While all detections made by SoFiA were visually inspected to discard obvious false positives caused by noise peaks or artefacts, a small number of false detections is likely to remain in the final catalogue. This number is expected to be at the 1\% level and will not affect large-scale statistical studies in any significant way. False positives are most likely to occur close to the detection threshold, and caution is therefore advised when using individual faint \ion{H}{I} detections in scientific studies (see comments in Table~\ref{tab_comments} of \ref{sec_comments}).

Another aspect to be aware of is that an individual \ion{H}{I} detection in the catalogue need not correspond to an individual galaxy. In a few cases, compact or interacting systems may have been detected as a single catalogue entry if their \ion{H}{I} emission is connected in phase space. Likewise, very faint objects near the detection threshold may have been detected only partially by SoFiA. Table~\ref{tab_comments} of \ref{sec_comments} contains a list of comments on individual detections that may have been merged or only partially detected.

\subsubsection{Hydra TR1 and TR2}
\label{sec_hydra_sf}

The Hydra field was subject to two blind source finding runs internally released to the WALLABY team. The catalogues from both runs are included in this public data release. The first run (Hydra TR1) was carried out on a smaller sub-region of the full field centred on the Hydra cluster itself. This sub-region is relatively clean and not as severely affected by continuum residuals and ODC ripples as the rest of the Hydra field, allowing us to extract a fairly deep source catalogue. A second run (Hydra TR2) on the full field was carried out with slightly modified SoFiA settings to limit the number of artefacts in the preliminary catalogue, but at the cost of a slightly reduced completeness at low SNR.

As a result, 119 out of the 148 sources from Hydra TR1 are also included in Hydra TR2, while the remaining 29 \ion{H}{I} sources from Hydra TR1 are missing from Hydra TR2 due to the slightly lower completeness of the SoFiA run on the full Hydra field. The significant overlap between Hydra TR1 and TR2 also means that only 592 of the 711~sources included in this release are actually separate \ion{H}{I} detections.

\subsubsection{Norma}

The western tile of the Norma field contains severe continuum artefacts from several bright sources. Unfortunately, as the artefacts extend across a significant fraction of the central region of the tile, we have been unable to extract a meaningful blind or targeted \ion{H}{I} catalogue from the western tile. The eastern tile of the Norma field, however, is not affected by strong continuum residuals, and we have been able to obtain a blind \ion{H}{I} catalogue similar in quality to the Hydra TR1 and TR2 catalogues.

\subsubsection{NGC~4636}
\label{sec_ngc4636}

Due to severe continuum artefacts from the bright quasar 3C~273 near the edge of the region, we were unable to carry out a blind \ion{H}{I} source finding run on either of the two tiles of the NGC~4636 field. We instead ran SoFiA locally on small sub-regions of the cube at the positions of previously detected optical and \ion{H}{I} sources with known redshifts (Lin et al., in prep.) using slightly modified SoFiA settings as listed in Table~\ref{tab_sofiapar}. For this purpose, galaxy positions and redshifts for the entire NGC~4636 field were extracted from DR16 of the Sloan Digital Sky Survey (SDSS; \citealt{Ahumada2020}), DR3 of the 6dF Galaxy Survey (6dFGS; \citealt{Jones2004,Jones2009}), the Cosmicflows--3 database \citep{Tully2016} and the Arecibo Legacy Fast ALFA Survey (ALFALFA; \citealt{Haynes2018}).

This strategy enabled us to extract the \ion{H}{I} emission from 147~galaxies across the NGC~4636 field, but at the expense of missing galaxies that are not included in any of the four redshift surveys used to select our target regions. In addition to this selection bias, several galaxies detected in the NGC~4636 field are still affected to some degree by strong continuum artefacts, and users are advised to exercise caution when using the NGC~4636 data in their scientific analysis.

\subsection{Source characterisation}
\label{sec_parameterisation}

In addition to source finding, SoFiA also provides measurements of fundamental source properties. A complete list of parameters included in the source catalogue is presented in Table~\ref{tab_catalogue} of \ref{app_data_products}. The calculation of some of the more complex source parameters is described in this section.

\subsubsection{Position and redshift}

Position and redshift in the source catalogue are derived from the flux-density-weighted centroid position, $(\mathcal{x}, \mathcal{y}, \mathcal{z})$, across the source mask in the original pixel coordinate system of the data cube:
\begin{equation}
    \mathcal{x} = \frac{1}{S_{\rm sum}} \sum_{i} \mathcal{x}_{i} S_{i}
\end{equation}
and likewise for $\mathcal{y}$ and $\mathcal{z}$, where $S_{i}$ is the flux density in pixel $i$, $S_{\rm sum} = \sum_{i} S_{i}$ is the summed flux density, and the summation is over all pixels of the source mask with positive flux density. The resulting pixel-based centroid is entirely arbitrary and therefore converted to more meaningful world coordinates (right ascension, declination and barycentric frequency) using WCS information from the FITS file header.

\begin{figure}[t!]
    \centering
    \includegraphics[width=0.8\linewidth]{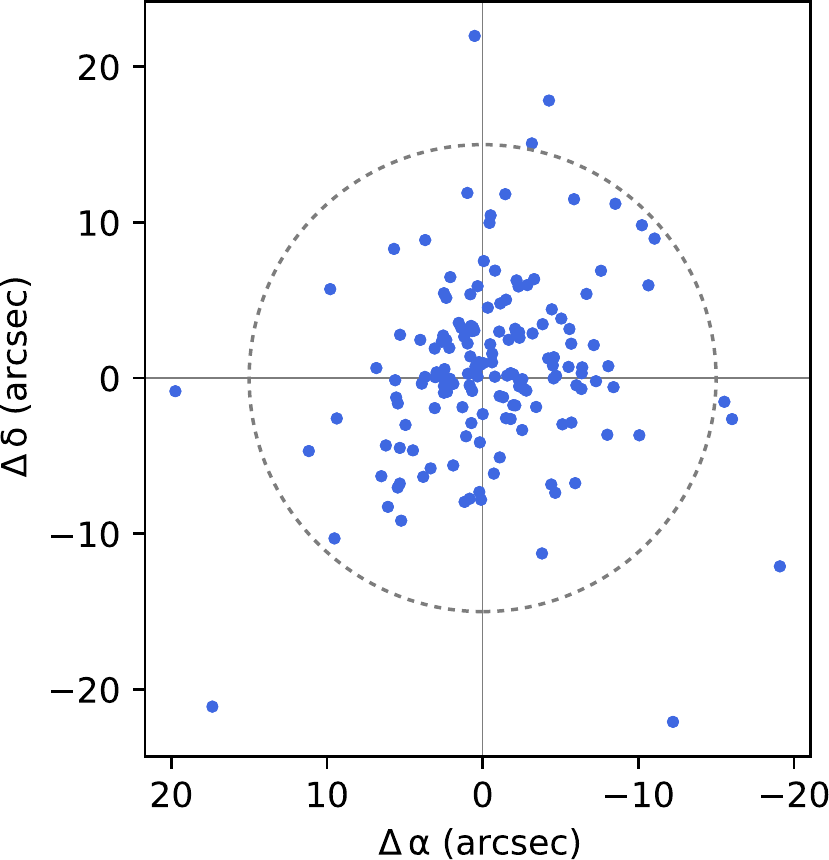}
    \caption{Position offset between 159~galaxies detected by WALLABY in the Hydra field and their optical counterparts retrieved from NED. The $30''$ synthesised beam size is indicated by the dashed circle.}
    \label{fig_position}
\end{figure}

Our position accuracy is illustrated in Fig.~\ref{fig_position} where we plot the relative position offset between the \ion{H}{I} detections from WALLABY and their optical counterparts retrieved from the NASA/IPAC Extragalactic Database (NED) for 159~galaxies from the Hydra field. An angular distance threshold of $30''$ and a maximum difference in redshift of $\pm 200~\mathrm{km \, s}^{-1}$ were applied to identify optical counterparts. Most of the resulting position offsets are much smaller than the $15''$ radius of WALLABY's synthesised beam (indicated by the dashed circle). The mean position offset of all galaxies within the beam radius in Fig.~\ref{fig_position} is $\langle \Delta \alpha \rangle = -0.5''$ (with an RMS of $4.3''$) and $\langle \Delta \delta \rangle = 0.6''$ (with an RMS of $4.6''$), which is consistent with zero and highlights the excellent recovery of sky positions by WALLABY.

The catalogued frequency, $\nu$, can be directly converted to barycentric redshift, $z$, via
\begin{equation}
    z = \frac{\nu_{0}}{\nu} - 1
\end{equation}
where $\nu_{0} = 1.42040575~\mathrm{GHz}$ is the rest frequency of the \ion{H}{I} 21-cm line transition. In the catalogue we provide basic Hubble distances, $\mathrm{c} z / H_{0}$ based on the barycentric redshift, assuming a local Hubble parameter of $H_{0} = 70~\mathrm{km \, s^{-1} \, Mpc^{-1}}$ \citep{Riess2016,Abbott2017,PlanckCollaboration2020}. However, given that the WALLABY pilot survey specifically targets nearby groups and clusters, many of the measured redshifts are likely to be dominated by peculiar motions. Hence, we caution against using the catalogued Hubble distances in any scientific analysis and instead recommend obtaining redshift-independent distances where possible.

We also provide statistical uncertainties of the pixel-based centroid position assuming Gaussian error propagation and additionally taking into consideration the spatial correlation of pixels due to the finite beam size:
\begin{equation}
    \sigma_{\mathcal{x}}^{2} = \frac{\Omega \sigma_{\rm rms}^{2}}{S_{\rm sum}^{2}} \sum_{i} (\mathcal{x}_{i} - \mathcal{x})^{2}
\end{equation}
and likewise for $\sigma_{\mathcal{y}}$ and $\sigma_{\mathcal{z}}$, where $\sigma_{\mathcal{x}}$ is the standard deviation in $\mathcal{x}$, $\Omega$ is the synthesised beam solid angle in units of pixels, and $\sigma_{\rm rms}$ is the local RMS noise level in the data cube. These uncertainties can be converted to physical units by multiplying by the spatial pixel size of $6''$ or spectral channel width of $1/54~\mathrm{MHz}$.

\begin{figure*}[!t]
    \centering
    \includegraphics[width=\linewidth]{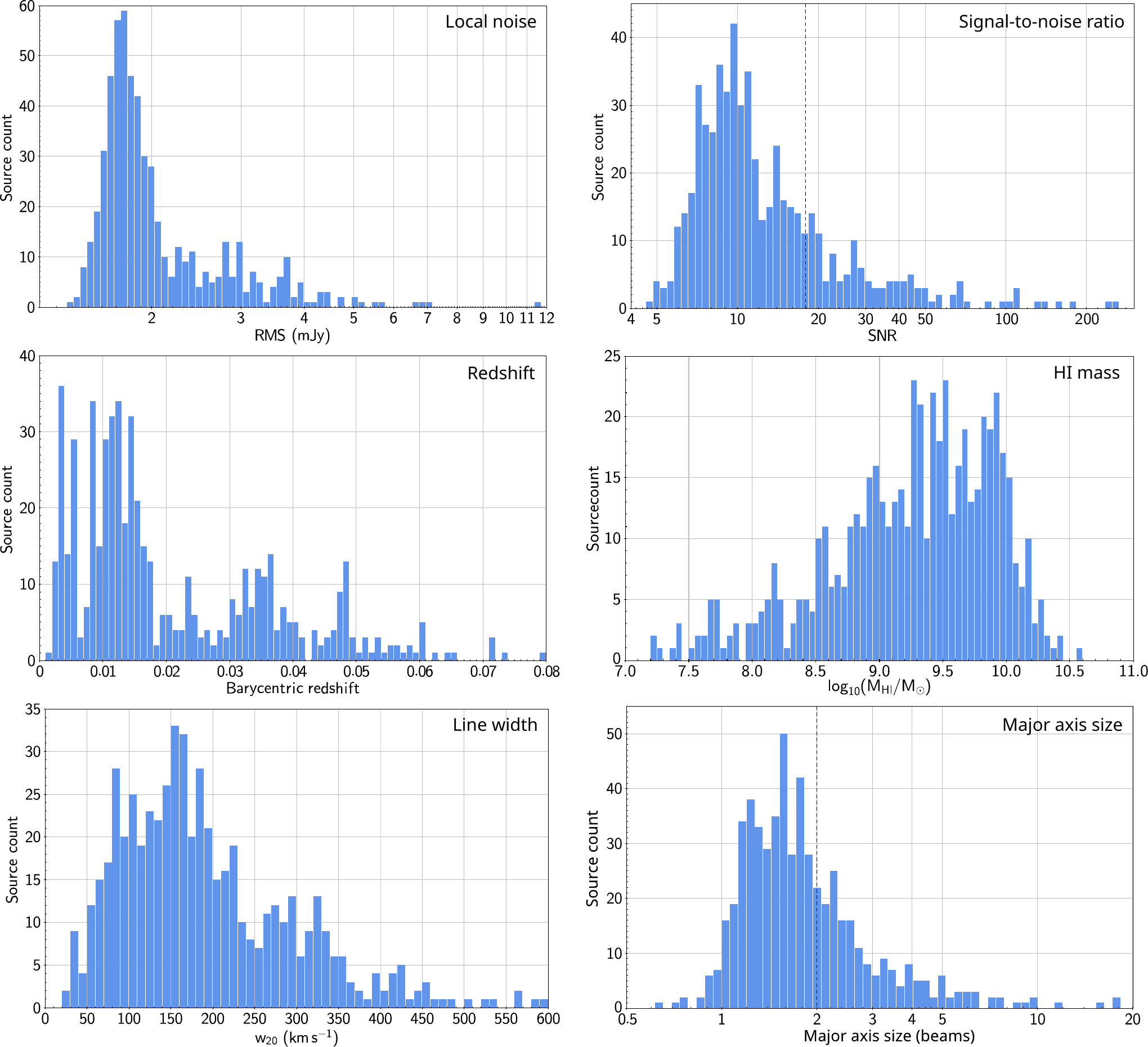}
    \caption{Histograms of local RMS noise level (top left), integrated signal-to-noise ratio (top right), barycentric redshift (centre left), \ion{H}{I} mass (centre right), $w_{20}$ line width in the source rest frame (bottom left) and \ion{H}{I} disc major axis size in units of the $30''$ synthesised beam size (bottom right) for all detections from Hydra~TR2, Norma~TR1 and NGC~4636~TR1. The dashed vertical lines in the top-right and bottom-right panels mark the selection thresholds for galaxies for which kinematic modelling was attempted (see Section~\ref{sec_kin}).}
    \label{fig_results}
\end{figure*}

\begin{figure}[!t]
    \centering
    \includegraphics[width=\linewidth]{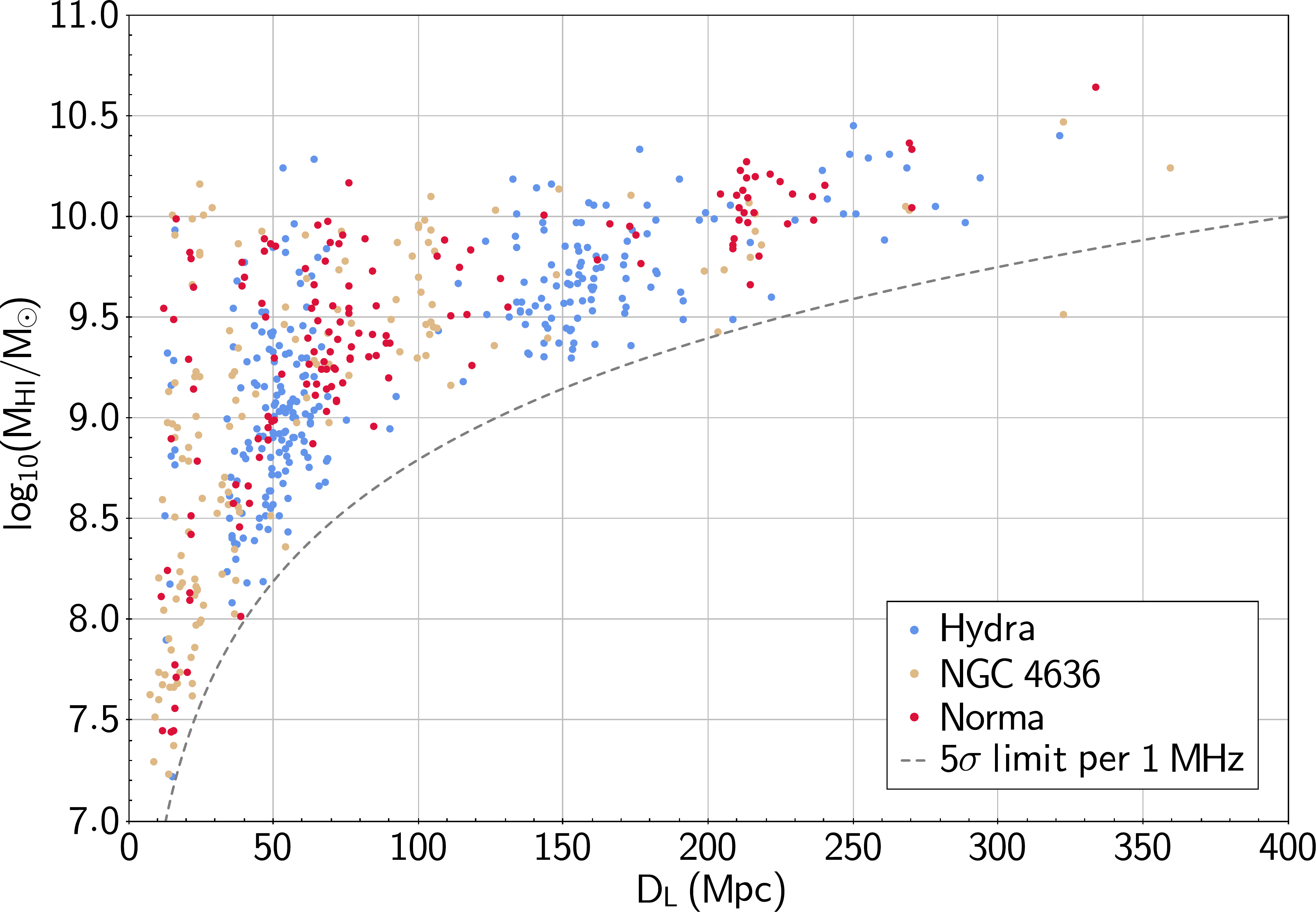}
    \caption{\ion{H}{I} mass plotted against luminosity distance, colour-coded by field, for the Hydra~TR2, NGC~4636~TR1 and Norma~TR1 data. The dashed, grey line indicates the $5 \sigma$ detection threshold for a point source across a frequency band of $1~\mathrm{MHz}$ (approximately $200~\mathrm{km \, s}^{-1}$ at $z = 0$) based on the median local RMS noise level of $1.85~\mathrm{mJy}$.}
    \label{fig_mass_dist}
\end{figure}

\subsubsection{Flux and \ion{H}{I} mass}

In addition to position and redshift, the integrated flux of a source is another critical parameter of any \ion{H}{I} survey, as it is needed to derive the \ion{H}{I}~mass of a galaxy. WALLABY fluxes are calculated as
\begin{equation}
    F = \frac{\Delta \nu S_{\rm sum}}{\Omega} \label{eqn_flux}
\end{equation}
where $\Delta \nu = 1/54~\mathrm{MHz}$ is the native channel width of WALLABY data which corresponds to a velocity width of approximately $4~\mathrm{km \, s}^{-1}$ at $z = 0$. According to Eq.~\ref{eqn_flux}, accurate flux measurements require knowledge of the beam solid angle, $\Omega$. All catalogued fluxes were calculated assuming the solid angle of the Gaussian restoring beam used during deconvolution,
\begin{equation}
    \Omega = \frac{\uppi \vartheta_{a} \vartheta_{b}}{4 \ln(2)}
\end{equation}
where $\vartheta_{a}$ and $\vartheta_{b}$ denote the FWHM of the major and minor axis of the restoring beam in units of pixels. In the case of WALLABY data, the restoring beam is very close to $30''$ in size, hence $\vartheta_{a} = \vartheta_{b} = 5$ for the $6''$ pixel size set in the imaging pipeline. This implies $\Omega = 28.3$ (in units of pixel area).

The accuracy of our flux measurement will be severely limited in situations where the \ion{H}{I} emission is not fully deconvolved, as the resulting beam will be a combination of the dirty beam and the restoring beam, and the solid angle of the beam will be ill-defined. We refer the reader to Section~\ref{sec_flux} where this issue is investigated and discussed in more detail.

We also provide an estimate of the statistical uncertainty, $\sigma_{F}$, of the flux measurement using Gaussian error propagation:
\begin{equation}
    \sigma_{F} = \sqrt{\frac{N}{\Omega}} \Delta \nu \sigma_{\rm rms}
\end{equation}
where $N$ is the total number of pixels contained within the 3D source mask. This again takes into account the degree of spatial correlation between pixels due to the synthesised beam size.

Lastly, fluxes from the WALLABY source catalogue can be converted to \ion{H}{I} masses using equation~48 from \citet{Meyer2017},
\begin{equation}
    \frac{M_{\ion{H}{I}}}{M_\sun} = 49.7 \left( \frac{D_{\rm L}}{\mathrm{Mpc}} \right)^{\! 2} \frac{F}{\mathrm{Jy \, Hz}} , \label{eqn_himass}
\end{equation}
under the additional assumption that the gas is optically thin. It should be noted that this conversion is non-trivial, as it depends on luminosity distance, $D_{\rm L}$ and hence on the choice of a cosmological model and peculiar velocity estimate. In the catalogue we provide a rough estimate of the \ion{H}{I} mass for each source based on the local, barycentric Hubble distance by assuming $D_{\rm L} = \mathrm{c} z / H_{0}$. Given the expected large systematic errors caused by peculiar motions, the catalogued \ion{H}{I} masses should only be considered as order-of-magnitude estimates, and users of the catalogue are strongly advised to derive their own \ion{H}{I} masses using Eq.~\ref{eqn_himass} in combination with more accurate distance measurements.

\subsubsection{Spectral line width}

The WALLABY catalogue provides two measures of spectral profile width, namely the widths at levels of 20\% and 50\% of the peak flux density in the integrated spectrum ($w_{20}$ and $w_{50}$, respectively). Both are measured by moving inwards from the edges of the spectrum until the signal is found to exceed the threshold for the first time. For improved accuracy we carry out a linear interpolation across the bracketing channels in between which the signal crosses the threshold. Note that the resulting line widths have not yet been corrected for the finite spectral resolution of the data, although the resulting error is entirely negligible for the broad lines found in most galaxies ($< 0.1\%$ for line widths of $> 100~\mathrm{km \, s}^{-1}$). We do not provide statistical uncertainties either at this point.

It is straightforward to convert the raw line width in units of frequency to source rest-frame velocity via
\begin{equation}
    \Delta v_{\rm src} = \frac{\mathrm{c} (1 + z)}{\nu_{0}} \Delta \nu_{\rm obs}
\end{equation}
\citep{Meyer2017} where $\mathrm{c}$ is the speed of light, $z$ is the cosmological redshift of the source, $\nu_{0} = 1.42040575~\mathrm{GHz}$ is the rest frequency of the \ion{H}{I} line and $\Delta \nu_{\rm obs}$ is the observed frequency width from the catalogue ($w_{20}$ or $w_{50}$).

The $w_{20}$ and $w_{50}$ measurements currently supplied in the WALLABY catalogue are rudimentary, as they are often affected by the stochastic noise in the data; $w_{50}$ in particular has been found to be generally too small, most notably in fainter sources, due to the peak flux density of the spectrum being biased by the noise \citep{Westmeier2021}. As we do provide the integrated spectrum of each source as part of this data release, we encourage users to derive their own line width measurements using more sophisticated methods with better accuracy, e.g.\ spectral profile fitting.

\subsubsection{Source size and orientation}

Source sizes and orientations are measured by fitting an ellipse to the integrated flux map (moment~0) of each source using the method presented in \citet{Banks1995}. The resulting major and minor axis sizes, $a$ and $b$, correspond to twice the dispersion along those axes. For example, if the surface brightness profile was described by an elliptical Gaussian of standard deviation $\sigma_{a}$ and $\sigma_{b}$, then $a = 2 \sigma_{a}$ and $b = 2 \sigma_{b}$.

We provide two different ellipse fits: a flux-weighted fit to all positive pixels of the flux map, and a non-weighted fit to all pixels in the flux map that have a flux of $F > 3 \sigma_{F}$, where $\sigma_{F} = \sigma_{\rm rms} \sqrt{N_{\rm chan}}$ is the statistical uncertainty of the flux based on the local RMS noise level, $\sigma_{\rm rms}$ and the number of spectral channels, $N_{\rm chan}$, across which the flux was summed. It should be noted that the resulting major and minor axis sizes and position angle have not been corrected for the effect of the synthesised beam size. Moreover, the position angle will be relative to the pixel grid of the data cube rather than the sky coordinate system.

\subsubsection{Kinematic major axis}

We also supply a basic estimate of the position angle of the kinematic major axis of each source on the sky, although we note that this value may not be meaningful for sources that do not show a noticeable velocity gradient, e.g.\ due to small angular size, face-on geometry or a genuine absence of rotation. The position angle is calculated by measuring the centroid of the \ion{H}{I} emission on the sky in each individual spectral channel across the source mask and then carrying out an orthogonal regression to the line of centroids on the sky. The resulting kinematic major axis position angle will point towards the approaching side of the object and is measured counter-clockwise, with $0\degr$ pointing up (in the direction of increasing $\mathcal{y}$). It is important to note that the position angle has not been corrected for the effects of beam smearing and will be relative to the pixel grid of the data cube, not the sky coordinate system. In addition, the kinematic major axis position angle measured by SoFiA might differ from the value derived by the kinematics pipeline \citep{Deg2022}.

\subsection{Quality and kinematic flags}

Each catalogue entry has two flags: a quality flag that indicates whether some of the emission from the source could be missing, and a kinematic flag that indicates whether kinematic modelling was attempted.

The following quality flag values are supported by SoFiA: 0 = no issues; 1 = source may be truncated along the spatial edge of the cube; 2 = source may be truncated along the spectral edge of the cube; 4 = source may be partially flagged. Flag values are additive, e.g.\ a value of $3 = 2 + 1$ indicates that the source may have been truncated along the spatial and spectral edge of the data cube at the same time. In principle, sources located on the edge of a cube were discarded, and only flag values of $0$ or $4$ can occur in the final catalogue.

The following kinematic flag values are supported: 0 = kinematic modelling was not attempted; 1 = kinematic modelling was attempted but was not successful; 2 = kinematic modelling was attempted and successful. A full description of the kinematic modelling approach and the definition of a successful model is given in \citet{Deg2022} and summarized in Section~\ref{sec_kin}.

\subsection{Source data products}

In addition to the source catalogue, we also provide basic data products for each individual object, including small cubelets, moment maps and integrated spectra. A summary of available data products is presented in Table~\ref{tab_products} of \ref{app_data_products}. These products can be used to carry out additional measurements of source parameters that are not currently provided in the default catalogue. An example set of data products for the bright galaxy WALLABY J165901$-$601241 in the Norma field is presented in Fig.~\ref{fig_example_output}, while Fig.~\ref{fig_example_output_2} shows the same products for the faint and distant galaxy WALLABY J101018$-$265209.

\subsection{Overview of basic galaxy properties}

In Fig.~\ref{fig_results} we present an overview of the basic properties of the detections, excluding Hydra TR1 sources, most of which are contained in Hydra~TR2.

The histogram of local RMS noise levels in the top-left panel of Fig.~\ref{fig_results} reveals a sharp peak at approximately $1.75~\mathrm{mJy}$, with a median value of $1.85~\mathrm{mJy}$. This is marginally higher than the nominal $1.6~\mathrm{mJy}$ of WALLABY. The elevated noise levels are readily explained by the effect of primary-beam attenuation along the edges of each field; the full survey will have a continuous sky coverage of spatially overlapping tiles and hence will not be affected by this issue. Moreover, the phase~1 pilot fields were specifically chosen to contain foreground groups and clusters at low redshifts where the system temperature, and hence the noise level, is intrinsically higher \citep{Hotan2021}.

In the top-right panel of Fig.~\ref{fig_results} we show a histogram of integrated signal-to-noise ratio defined as the integrated flux divided by its statistical uncertainty. The distribution peaks at $\text{SNR} \approx 8$--$9$, which is somewhat higher than the threshold of $5$ anticipated for the full WALLABY survey. The presence of artefacts in the data, most notably residual continuum emission and the ODC ripple, did not allow for the source finder to be pushed to its theoretical limit during phase~1 of WALLABY pilot observations (see Section~\ref{sec_data_quality}). However, these issues are expected to be rectified before the start of full survey observations to enable better completeness down to $\text{SNR} \approx 5$.

The distribution of \ion{H}{I} detections in redshift space is shown in the centre-left panel of Fig.~\ref{fig_results}. Most pilot survey detections are found at low redshifts of $z \lesssim 0.02$, with only few galaxies detected near the high-redshift end of the RFI-free frequency band accessible to WALLABY. This distribution is the result of a selection effect, as the pilot survey fields were specifically chosen to contain rich group and cluster environments in the local Universe, but happened to contain not much structure in the background; the redshift distribution of the full WALLABY survey is expected to peak at higher redshift of $z \approx 0.04$ instead.

The centre-right panel of Fig.~\ref{fig_results} shows the \ion{H}{I} mass distribution of the detections. \ion{H}{I} masses were estimated from the raw flux measurements and barycentric redshifts under the assumption of a distance of $\mathrm{c} z / H_{0}$ with a local Hubble parameter of $H_{0} = 70~\mathrm{km \, s^{-1} \, Mpc^{-1}}$. As expected for a blind survey, the distribution peaks near $M_{\rm \ion{H}{I}}^{\ast} \approx 6 \times 10^{9}~M_{\sun}$ \citep{Zwaan2005}, with a sharp drop at the high-mass end caused by the genuine rarity of galaxies with $M_{\rm \ion{H}{I}} > 10^{10}~M_{\sun}$ and an extended tail at the low end which is due to the limited redshift range (and hence volume) across which low-mass galaxies are in principle detectable.

Histograms of $w_{20}$ line width (in the source rest frame) and \ion{H}{I} disc major axis size (defined as two times the profile dispersion, $2 \sigma$, along the major axis divided by the beam FWHM) are presented in the two bottom panels of Fig.~\ref{fig_results}. While WALLABY detections are generally well-resolved in the spectral domain, the vast majority of galaxies detected by WALLABY is only marginally resolved by the $30''$ synthesised beam size. 69\% of all detections have a major axis size of less than 2~beams, while only 12\% are more than 3~beams in size. This makes WALLABY data ideal for large-scale statistical and cosmological studies, but will limit the ability to carry out detailed morphological studies of galaxies. Nevertheless, the full survey is expected to detect several thousand spatially well-resolved galaxies across the southern sky. In addition, we intend to image several thousand selected galaxies from the full WALLABY survey at a higher angular resolution of up to $10''$ to enable more detailed studies of their interstellar medium.

The resulting \ion{H}{I} mass plot as a function of luminosity distance in a flat $\Lambda$CDM Universe with $H_{0} = 70~\mathrm{km \, s^{-1} \, Mpc^{-1}}$ and a local matter density of  $\Omega_{\rm m} = 0.3$ is shown in Fig.~\ref{fig_mass_dist}, colour-coded by field. The theoretical detection limit for a $5 \sigma$ signal across a bandwidth of $1~\mathrm{MHz}$ based on the median local RMS noise level of $1.85~\mathrm{mJy}$ is shown as the dashed, grey line. Several large-scale structures are discernible, most notably the Hydra cluster at $D_{\rm L} \approx 60~\mathrm{Mpc}$. The plot illustrates that WALLABY is a shallow survey with a detection threshold that exceeds $M_{\rm \ion{H}{I}}^{\ast}$ at the far end of the redshift range processed here ($z \lesssim 0.085$ corresponding to $D_{\rm L} \approx 400~\mathrm{Mpc}$).

\section{Flux accuracy}
\label{sec_flux}

\begin{figure*}[!t]
    \centering
    \includegraphics[width=0.7\linewidth]{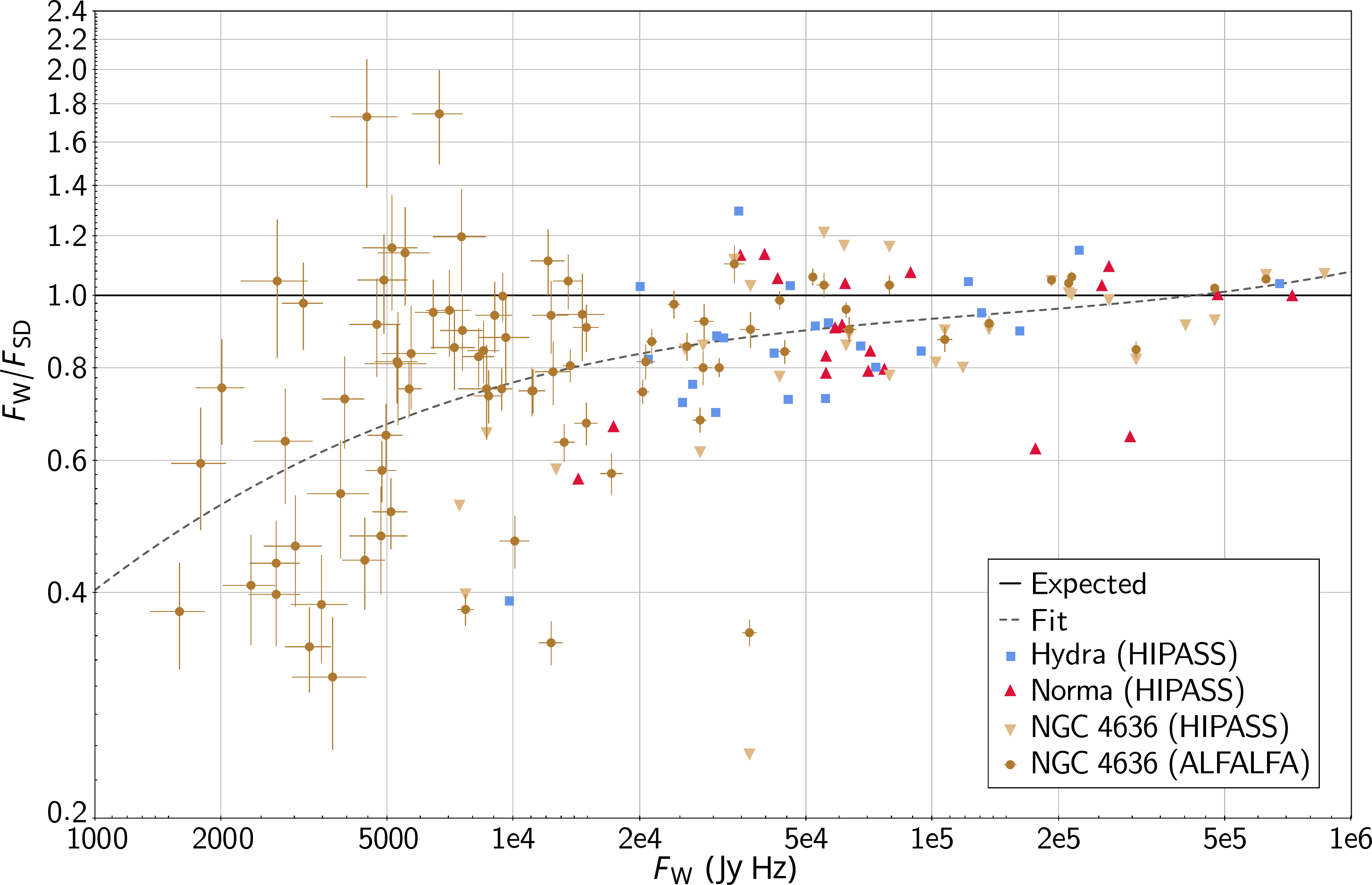}
    \includegraphics[width=0.7\linewidth]{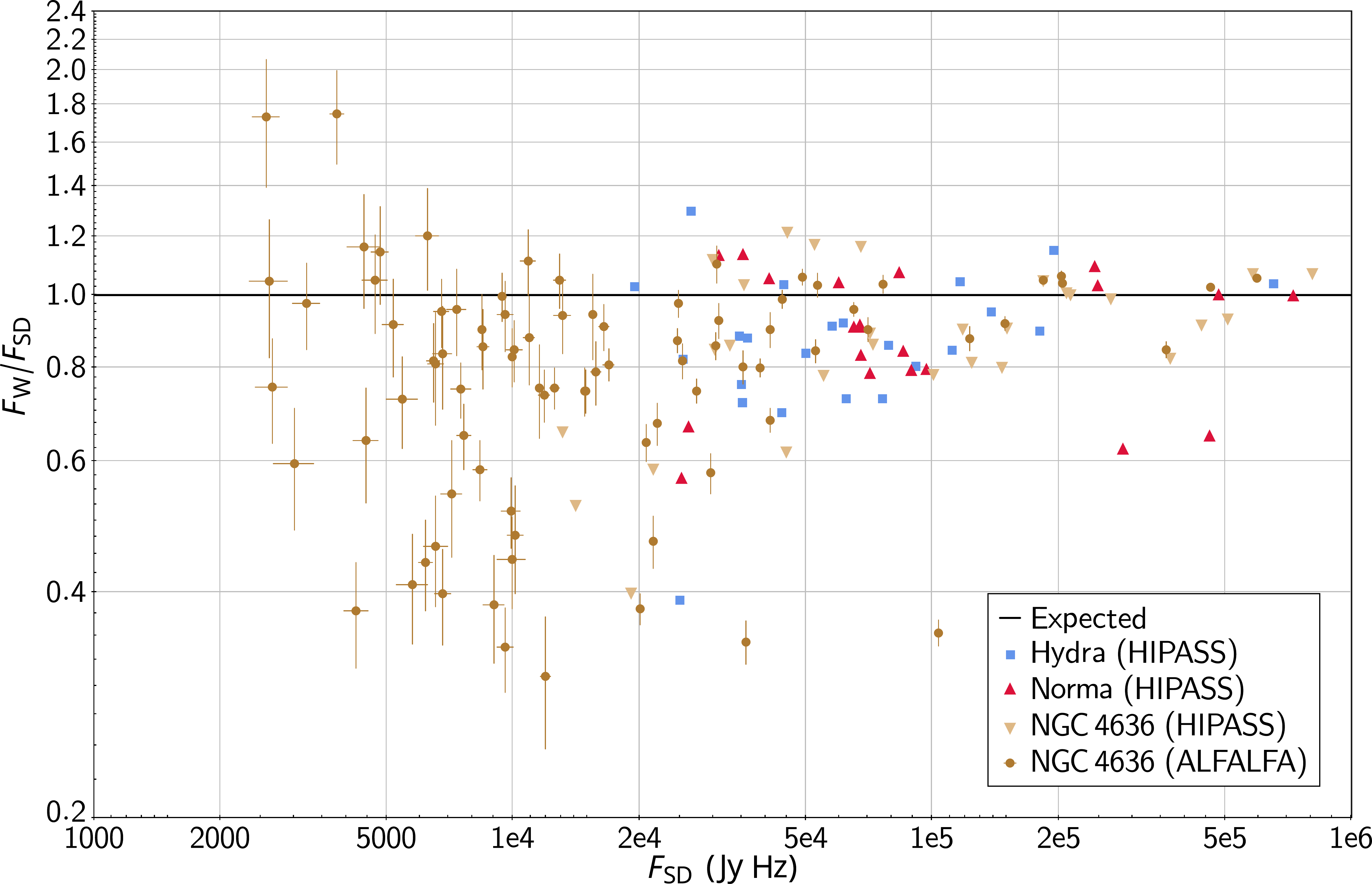}
    \caption{Ratio of WALLABY flux, $F_{\rm W}$, and single-dish flux from ALFALFA or HIPASS, $F_{\rm SD}$, as a function of WALLABY flux (top) and single-dish flux (bottom) for jointly detected galaxies in the three WALLABY pilot survey fields. The dashed, grey curve in the upper panel shows the result of a third-order polynomial fit to the data points in double-logarithmic space. Note that statistical uncertainties are only shown for ALFALFA data, as the HIPASS catalogue does not list individual flux measurement uncertainties.}
    \label{fig_flux_comparison}
\end{figure*}

One of the most critical parameters to be measured from WALLABY data is the total flux of the detected galaxies. In order to understand the accuracy of our flux measurement, we are comparing in Fig.~\ref{fig_flux_comparison} the ratio of the total flux measured by WALLABY, $F_{\rm W}$, and the corresponding flux from the ALFALFA and HIPASS single-dish surveys, $F_{\rm SD}$, for all galaxies from our pilot survey fields that were detected by either of those two surveys. It should be noted that ALFALFA is significantly deeper than HIPASS and therefore dominates the lower end of the flux range covered by WALLABY. At the same time, the NGC~4636 field is the only one containing ALFALFA detections, as the other two fields are located too far south and are outside of the ALFALFA sky coverage.

As shown in the top panel of Fig.~\ref{fig_flux_comparison}, WALLABY is accurately recovering the total flux of bright galaxies. However, there appears to be an increasing flux deficit for fainter sources, with less than half of the ALFALFA flux detected by WALLABY for some galaxies near the faint end of the flux range. This sharp drop is unexpected, as one would na\"{i}vely expect the flux of very faint galaxies near the detection threshold to be overestimated due to noise bias, as faint sources would remain undetected if the overall contribution of stochastic noise to the total flux measurement was negative.

It can be demonstrated that the observed decline in flux is in fact a perceived bias resulting from the fact that the two axes in the upper panel of Fig.~\ref{fig_flux_comparison} are coupled. It is therefore not possible to shift a data point upwards without simultaneously moving it to the right. As a result, the presence of statistical uncertainties in our flux measurement will inevitably create the perception of a declining flux at the faint end, because sources that we perceive as faint may not actually be faint, but could simply have the largest negative flux measurement errors.

This perceived bias can be demonstrated by plotting flux ratio against single-dish flux instead of WALLABY flux, as shown in the bottom panel of Fig.~\ref{fig_flux_comparison}. Indeed, the flux ratio now rises near the faint end. It is obvious, though, that the WALLABY fluxes on average are too low across almost the entire flux range with the exception of the upper end. The most likely explanation for this significant flux deficit is insufficient deconvolution of fainter sources. The ASKAP data reduction pipeline currently deconvolves each primary beam of each footprint separately, which limits the depth to which the data can be deconvolved, in particular compared to the final noise level after mosaicking of the 36~beams into a single footprint and combination of the two interleaved 8-hour footprints into the final tile. It should be noted that other deconvolution-related issues have the potential to additionally contribute to this problem, e.g.\ the use of sub-optimal deconvolution thresholds or masks.

As a result of this procedure and the fact that most WALLABY detections are faint to the point of sitting below the noise level in an individual frequency channel, it is likely that a significant fraction of WALLABY detections was only partially deconvolved. Due to the solid angle of the dirty beam being ill-defined, it is therefore difficult to characterise the flux of faint sources in our data, in particular for those that are spatially extended. In principle, the deconvolution-related flux deficit can be reduced by deconvolving to lower flux levels within tight source masks. However, this would require the deconvolution step and subsequent mosaicking of footprints and tiles by the data reduction pipeline to be repeated after source finding. Unfortunately, the ASKAP data reduction workflow is not currently set up to provide such functionality.

In the interim, a straightforward method of improving the accuracy of fluxes measured in WALLABY pilot survey data would be to characterise the flux deficit and apply a statistical correction. Under the assumption that the single-dish fluxes are accurate, we fit a third-order polynomial function to the data points in the upper panel of Fig.~\ref{fig_flux_comparison}. The resulting best fit yields
\begin{equation}
    \log_{10} \! \left( \frac{F_{\rm W}}{F_{\rm SD}} \right) = 0.0285 \varphi^{3} - 0.439 \varphi^{2} + 2.294 \varphi - 4.097
\end{equation}
where
\begin{equation}
    \varphi = \log_{10} \! \left( \frac{F_{\rm W}}{\mathrm{Jy \, Hz}} \right).
\end{equation}
The polynomial fit is shown as the dashed, grey line in the upper panel of Fig.~\ref{fig_flux_comparison} and can be used to statistically correct all fluxes in the WALLABY phase~1 pilot survey catalogue by dividing them by the polynomial fit (using the logarithmic quantities).

It should be noted that a significant scatter will remain, as the precise flux deficit will depend not primarily on the integrated flux, but on a range of different metrics such as source size and morphology, actual shape of the dirty beam, fractional deconvolution, etc. The overall standard deviation in the upper panel of Fig.~\ref{fig_flux_comparison} after division by the polynomial fit is approximately $0.126~\mathrm{dex}$ which corresponds to a scatter of about 34\%. This value does vary somewhat with flux, with faint sources experiencing a larger scatter than bright ones.

While it is not entirely clear how much of this scatter is caused by flux errors in ALFALFA, potential users of WALLABY pilot survey data will need to keep in mind that fluxes in the phase~1 pilot catalogue are uncertain to within several ten percent even after applying the statistical correction presented here, in particular for fainter sources.

\section{Kinematic models}
\label{sec_kin}

The bottom-right panel of Fig.~\ref{fig_results} shows that the vast majority of the pilot survey detections are spatially resolved, albeit only marginally so. This is representative of the full WALLABY survey, which will spatially resolve thousands of detections \citep{Koribalski2020}; a custom-built pipeline that is optimised for kinematic modelling in this marginally-resolved regime is being developed. 

In the interim, for the pilot survey data we automatically apply rotating, flat-disk 3D tilted-ring models using Fully Automated TiRiFiC (FAT; \citealt{Kamphuis2015}) and 3D-Barolo \citep{diTeodoro15} to suitable detections, and take the average of their outputs to produce an optimised model and estimate uncertainties. The technique produces homogeneously-derived kinematic model geometries, rotation curves and surface density distributions for \nKin{} spatially-resolved sources, \nKinUnique{} of which are unique, while the remaining \nKinOverlap{} galaxies are included in both Hydra~TR1 and TR2 and thus have been modelled twice.  A full description of the modelling technique, outputs, kinematic sample properties and associated data release is provided in a companion paper \citep{Deg2022}. Here, we give a brief overview of the approach.

\begin{figure*}[!t]
    \centering
    \includegraphics[width=0.9\linewidth]{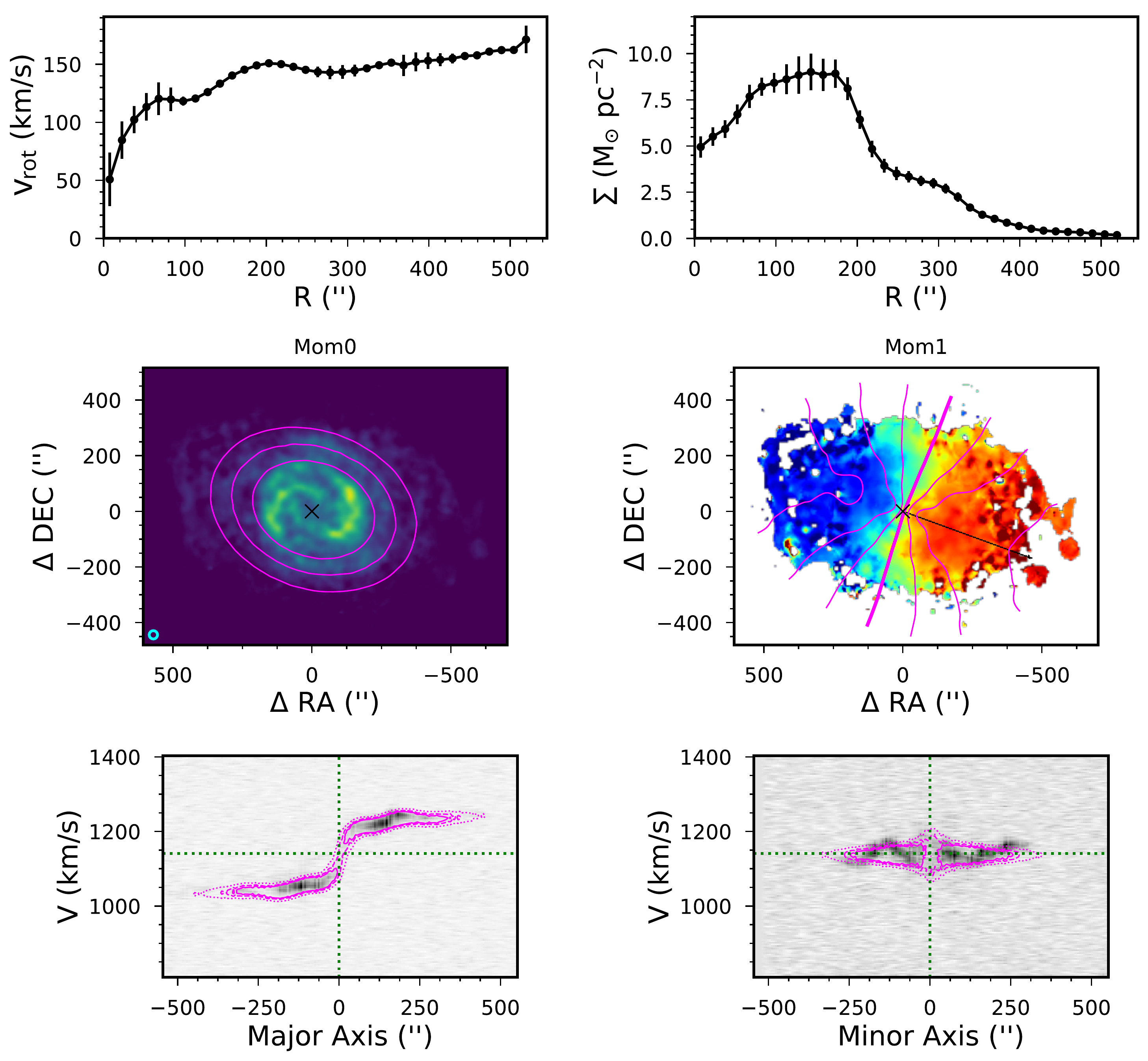}
    \caption{Example output products from 3D kinematic modelling of WALLABY J165901$-$601241, for which the SoFiA output products are shown in Fig.~\ref{fig_example_output}. The top row shows the rotation curve (left) and projected surface density profile (right) of the optimised axisymmetric rotating-disk model. The middle row shows ellipses with the optimised kinematic disk geometry overplotted on the $0^{\rm th}$ spectral moment (left), and the projected optimised model isovelocity contours overplotted on the $1^{\rm st}$ spectral moment (right). The circle in the corner of the middle-left panel shows the restoring beam FWHM. The thick solid line in the middle-right panel shows the minor axis of the optimised model, the black arrow shows the kinematic position angle and the black cross shows the kinematic centre. The bottom row shows isodensity contours of the optimised axisymmetric model overplotted on major-axis (left) and minor-axis (right) position-velocity slices through the data, with the lowest model contour set to five times the noise in the slice. In both panels of the bottom row, the dotted green lines cross at the kinematic centre of the optimised model.}
    \label{fig_kin_summary}
\end{figure*}

We attempt to kinematically model all 209/592 source detections for which the number of major axis beams (bottom-left panel in Fig.~\ref{fig_results}) exceeds two, or for which the integrated signal-to-noise (top-right panel in Fig.~\ref{fig_results}) satisfies $\log(\mathrm{SNR}) \geq 1.25$, regardless of the detection morphology. For reference, the detection of WALLABY~J165901$-$601241 in Fig.~\ref{fig_example_output} well exceeds both criteria, while that for WALLABY~J101018$-$265209 in Fig.~\ref{fig_example_output_2} satisfies neither one.

For each source that meets the size and SNR criteria, we Hanning-smooth the SoFiA cubelet by three spectral channels to increase the per-pixel signal-to-noise ratio, which we find improves model stability. We then automatically apply rotating, flat-disk 3D models to each cubelet using both FAT and 3D-Barolo in the same manner across all selected sources. The two algorithms differ significantly in initialisation and optimisation even when applying the same underlying model to a dataset \citep{diTeodoro15, Kamphuis2015}. We therefore use differences between the code outputs to judge model reliability and estimate uncertainties.

We visually examine the output models to assess whether or not they are successful, and only sources for which both 3D-Barolo and FAT produce physically plausible models are further considered. For those sources, we average the best-fitting disk geometries of the successful model pairs to derive the optimised disk geometry. We then derive an interpolated average rotation curve from the individual model outputs, inclination-corrected to the optimised disk geometry, and adopt half the difference between the two models as the uncertainty. A projected surface density profile is then derived from isodensity fits to the SoFiA moment~0 map using the optimised disk geometry. Finally, model datacubes and moments in formats matching their spectrally smoothed and full-resolution SoFiA cubelet counterparts are generated using the MCGSuite package.\footnote{\url{https://github.com/CIRADA-tools/MCGSuite}}

In all, we produce kinematic models for \nKin spatially-resolved detections across the three pilot survey fields: 35 in Hydra, 31 in Norma, and 43 in NGC~4636. For each successful model, the optimised disk geometry (centre, systemic velocity, position angle, inclination), rotation curve, projected surface density distribution and their associated uncertainties, as well as model datacubes, are publicly released (see Section~\ref{sec_data_access}).  A full description of all products is given in \citet{Deg2022}.

Fig.~\ref{fig_kin_summary} shows an example of the kinematic models for WALLABY~J165901$-$601241, for which the source finding output products are shown in Fig.~\ref{fig_example_output}. The optimal rotation curve and projected surface density distribution are in the top left and right panels, where adjacent points are separated by half a restoring beam FWHM. In the left and right panels of the second row, concentric ellipses with the kinematic disk geometry and projected isovelocity contours from the optimised kinematic model are overplotted on the SoFiA moment~0 and moment~1 maps, respectively. The bottom row of Fig.~\ref{fig_kin_summary} shows position-velocity slices through the cubelet along the optimal kinematic major (left) and minor (right) axes, with model contours overplotted.

Fig.~\ref{fig_kin_summary} illustrates that the optimised model obtained is a good representation of WALLABY~J165901$-$601241, although there are complexities in its \ion{H}{I} distribution that have no counterparts in the rotating, flat-disk model applied. This is to be expected since the homogeneous modelling approach adopted targets the marginally spatially-resolved regime in which the majority of pilot survey detections lie. This detection is unusual for its high spatial resolution across the disk (c.f.~the bottom-right panel of Fig.~\ref{fig_results}); it is one of the few pilot survey detections which warrant detailed velocity map modelling (e.g.~\citealt{Oh2018}), which will be explored in future work.  

\begin{figure}[!t]
    \centering
    \includegraphics[width=0.9\linewidth]{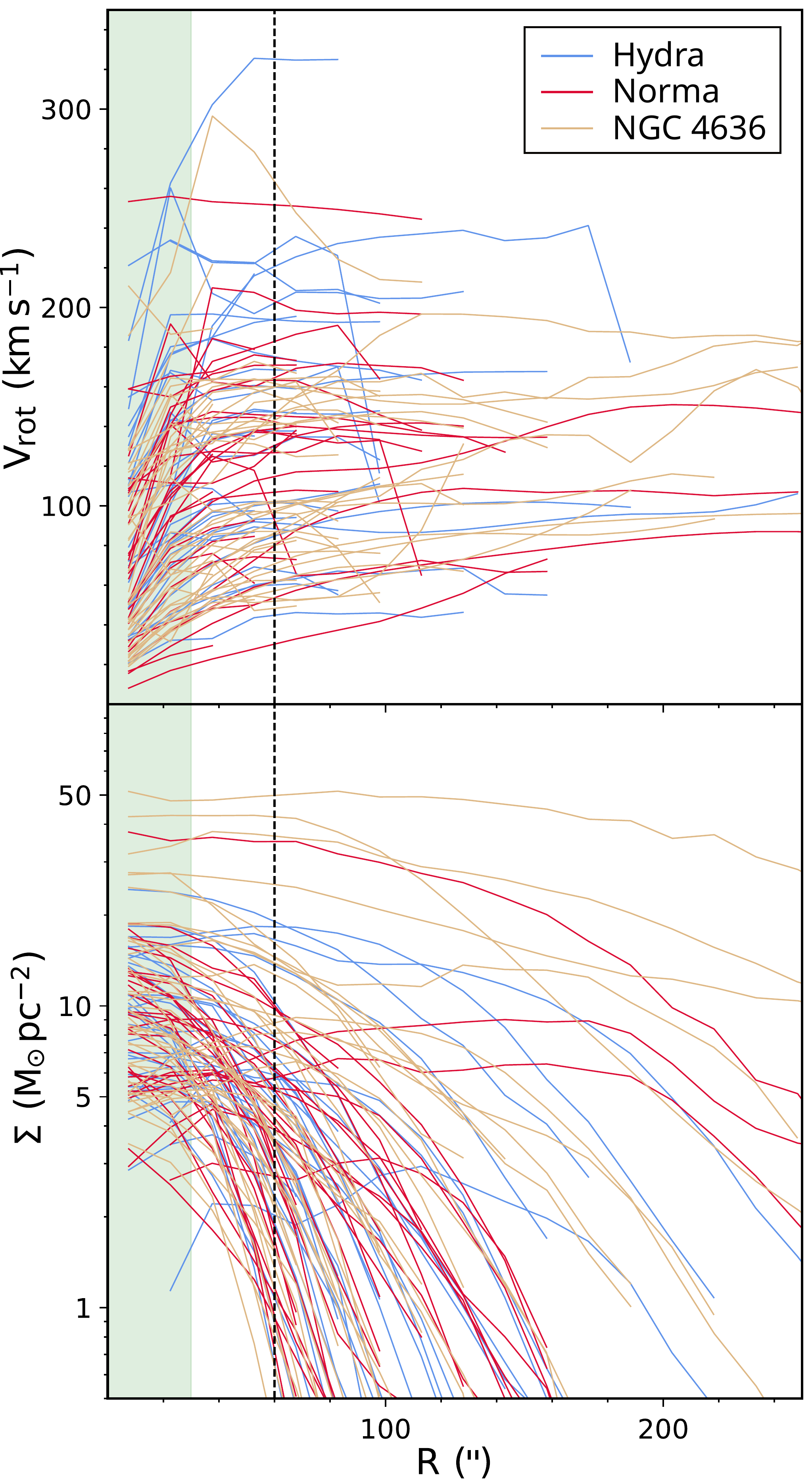}
    \caption{Distribution of optimised model rotation curves (top) and projected surface density distributions (bottom) across the three pilot fields. In both panels, the blue, red and orange lines correspond to galaxies from the Hydra, Norma and NGC~4636 fields, respectively. The light-green shaded region and vertical dashed line have radial extents equal to one ($= 30''$) and two ($= 60''$) times the restoring beam FWHM, respectively; the model values within the light-green region are likely impacted by beam smearing.}
    \label{fig_RCs+SBs}
\end{figure}

Fig.~\ref{fig_RCs+SBs} shows the rotation curves (top) and projected surface density distributions (bottom) of all the successful pilot survey source models in the Hydra (blue lines), Norma (red lines) and NGC~4636 (orange lines) fields. In both panels, the light-green shaded region and vertical dashed line denote radial extents of one ($= 30''$) and two ($= 60''$) times the restoring beam FWHM, respectively. Model values within the light-green shaded region are likely to be impacted by beam smearing; we recommend against their use in scientific applications in the absence of an estimate of the magnitude of this effect.

As expected from Fig.~\ref{fig_results}, Fig.~\ref{fig_RCs+SBs} shows that most modelled detections are only marginally spatially resolved, with only 11/\nKinUnique{} model rotation curves and 14/\nKinUnique{} model surface density distributions extending further than $200''$ ($\approx 6.5$ times the restoring beam FWHM).\footnote{The discrepancy in the numbers is caused by the fact that surface density profiles are measured beyond the outermost point of the rotation curve.} The variety of rotation curve and surface density distribution extents, amplitudes and shapes evident in Fig.~\ref{fig_RCs+SBs} is driven by a combination of extrinsic (e.g.~distance, inclination, resolution) and intrinsic (e.g.~mass distribution, dynamical state) factors. The physical properties of the population of the kinematically-modelled pilot survey galaxies, as well as the factors that dictate their modellability, are discussed in \citet{Deg2022}. 

\section{Data access}
\label{sec_data_access}

The WALLABY source catalogue, kinematic model catalogue and associated data products (e.g.\ cubelets, moment maps, integrated spectra, rotation curves, radial surface density profiles) are available online through the CSIRO ASKAP Science Data Archive (CASDA)\footnote{\url{https://research.csiro.au/casda/}} and the Canadian Astronomy Data Centre (CADC).\footnote{\url{https://www.cadc-ccda.hia-iha.nrc-cnrc.gc.ca/}} All source and kinematic model data products are made available at both locations. Links to the data access services and the software tools used to produce the data products as well as documented instructions and example scripts for accessing the data are available from the WALLABY Data Portal on the WALLABY website.\footnote{\url{https://wallaby-survey.org/data/}}

\subsection{Access via CASDA}

CASDA requires users to register and create an account to download data. Through the CASDA Data Access Portal (DAP) it is possible to download the entire catalogue and associated data products by searching for the corresponding collections (full data cubes: \url{https://doi.org/10.25919/xr39-jh63}; source catalogue and data products: \url{https://doi.org/10.25919/09yg-d529}; kinematic models: \href{https://doi.org/10.25919/213m-p819}{https://doi.org/10.25919/213m-p819}). The source finding and kinematics modelling products are available as separate collections in CASDA. The CASDA Observation Search\footnote{\url{https://data.csiro.au/domain/casdaObservation/}} can be used to search for and return catalogued source properties and associated data products for single or multiple objects, while the CASDA Skymap Search\footnote{\url{https://data.csiro.au/domain/casdaSkymap/}} uses an \emph{Aladin Lite} interface to search for and view the locations of sources on the sky. A user can then interactively click on individual sources of interest to view the source finding or kinematic modelling properties and download the individual source data products.

CASDA also offers Virtual Observatory (VO) services for querying catalogues. For example, using CASDA's Table Access Protocol (TAP) service within TOPCAT, the WALLABY catalogues are available under the ASKAP project ID for the WALLABY pilot survey (AS102). Using VO services to access the catalogue from CASDA does not require a user account.

CASDA also provides access to the full 30~deg$^2$ mosaicked spectral line cubes (72~beams of footprints A and B combined) from which the source catalogue was produced. These are available as a separate collection in CASDA through the DAP user interface and VO services. However, we caution the user against downloading the full cubes, as they are approximately $1~\mathrm{TB}$ in size. Using an account on CASDA, the user can instead create cutouts of smaller regions, either within the CASDA DAP user interface or by using the Simple Image Access Protocol (SIAP) to `discover' relevant files and the Server-side Operations for Data Access (SODA) protocol through a Python script or Jupyter notebook to select the region and channel range of interest. The CASDA module of the Astropy Astroquery package\footnote{\url{https://astroquery.readthedocs.io/en/latest/casda/casda.html}} can also be used to generate a cutout of any cubes of interest.

\subsection{Access via CADC}

CADC also offers Virtual Observatory services for querying catalogues and data products, but unlike CASDA does not require user registration. CADC provides user guides for using TAP services\footnote{\url{https://www.cadc-ccda.hia-iha.nrc-cnrc.gc.ca/en/doc/tap/}} and ADQL queries.\footnote{\url{https://github.com/opencadc/notebook-tutorials}}

The CADC Advanced Search tool\footnote{\url{https://www.cadc-ccda.hia-iha.nrc-cnrc.gc.ca/en/search/}. Instructions on using the CADC Advanced Search are available at \url{https://www.cadc-ccda.hia-iha.nrc-cnrc.gc.ca/en/doc/advancedsearch/}.} is a browser interface to all of the multi-wavelength, multi-facility data collections that it stores. The interface itself builds ADQL queries based on the inputs to the search engine that can be used elsewhere. CADC Advanced Search is also integrated with a powerful `sky view' that shows the coverage of a particular area by WALLABY and other surveys in the archive.

\section{Summary}
\label{sec_summary}

We present WALLABY pilot data release~1 (PDR1), the first public release of \ion{H}{I} data from phase~1 of the WALLABY pilot survey on ASKAP, including a source catalogue, images and spectra for almost 600~galaxies detected in \ion{H}{I} emission as well as kinematic models of more than 100 spatially resolved galaxies ($> 2$~beams across). The data were taken in three separate $60~\mathrm{deg}^{2}$ fields on the sky in the direction of the Hydra and Norma galaxy clusters and the NGC~4636 galaxy group, although the western half of the Norma field is severely affected by residual radio continuum emission and for that reason omitted from this release.

While the noise varies significantly across each field and frequency band, we are generally close to achieving the target noise level of WALLABY of $1.6~\mathrm{mJy}$ per $30''$ beam and $18.5~\mathrm{kHz}$ spectral channel throughout most of the cosmological volume covered by the observations. The bandwidth of 144~MHz enables \ion{H}{I} emission to be detected out to a redshift of $z \approx 0.08$, corresponding to a maximum recession velocity of $\mathrm{c} z \approx 24,000~\mathrm{km \, s}^{-1}$. WALLABY's target sensitivity translates into a $5 \sigma$ \ion{H}{I} mass sensitivity for point sources of about $5.2 \times 10^{8} \, (D_{\rm L} / 100~\mathrm{Mpc})^{2} \, M_{\sun}$ over 50 spectral channels ($\approx 200~\mathrm{km \, s}^{-1}$) and a $5 \sigma$ \ion{H}{I} column density sensitivity of about $8.6 \times 10^{19} \, (1 + z)^{4} \, \mathrm{cm}^{-2}$ over 5 spectral channels ($\approx 20~\mathrm{km \, s}^{-1}$) for diffuse emission filling the $30''$ synthesised beam.

Being a pilot survey, the current data are still affected by a number of technical issues and artefacts, including deconvolution artefacts, residual sidelobes, residual continuum emission, bandpass ripples caused by the ASKAP on-dish calibrators, correlator drop-outs and uncertainties in relation to the precise shape and size of the primary beams supplied by ASKAP's phased array feeds. Users of the data need to be aware that these issues will have some impact on the quality of the data, most notably in the form of systematic flux errors of up to several 10\% depending on the exact location and flux density of the source. Most of these remaining problems and uncertainties are expected to be rectified by the start of full WALLABY survey observations.

The final source catalogue may contain a small number of false detections (at the 1\% level), in particular at the faint end of the flux range where genuine \ion{H}{I} signals become difficult to separate from artefacts. In other cases, faint galaxies may have been only partially detected, while close galaxy pairs or interacting systems may have been merged into a single \ion{H}{I} detection. Table~\ref{tab_comments} of \ref{sec_comments} contains comments on individual detections that may have been affected by these or other issues.

Users should also note that most of the NGC~4636 field is severely affected by strong continuum residuals which did not allow us to carry out a blind source finding run on that field. Instead, \ion{H}{I} detections in the NGC~4636 field were extracted by running SoFiA on small subregions around the positions and redshifts of known galaxies obtained from existing optical and \ion{H}{I} redshift catalogues. As a result, the selection biases for the NGC~4636 field are rather different from the other two fields which were subjected to fully blind \ion{H}{I} source finding. In addition, some of the detections in the NGC~4636 field could be affected by continuum artefacts.

\begin{acknowledgement}
    The Australian SKA Pathfinder is part of the Australia Telescope National Facility (\url{https://ror.org/05qajvd42}) which is managed by CSIRO. Operation of ASKAP is funded by the Australian Government with support from the National Collaborative Research Infrastructure Strategy. ASKAP uses the resources of the Pawsey Supercomputing Centre. Establishment of ASKAP, the Murchison Radio-astronomy Observatory and the Pawsey Supercomputing Centre are initiatives of the Australian Government, with support from the Government of Western Australia and the Science and Industry Endowment Fund. We acknowledge the Wajarri Yamatji people as the traditional owners of the Observatory site.
    
    The Canadian Initiative for Radio Astronomy Data Analysis (CIRADA) is funded by a grant from the Canada Foundation for Innovation 2017 Innovation Fund (Project 35999) and by the Provinces of Ontario, British Columbia, Alberta, Manitoba and Quebec, in collaboration with the National Research Council of Canada, the US National Radio Astronomy Observatory and Australia's Commonwealth Scientific and Industrial Research Organisation.
    
    This research used the facilities of the Canadian Astronomy Data Centre operated by the National Research Council of Canada with the support of the Canadian Space Agency.
    
    This paper includes archived data obtained through the CSIRO ASKAP Science Data Archive, CASDA (\url{http://data.csiro.au}).
    
    WALLABY acknowledges technical support from the Australian SKA Regional Centre (AusSRC) and Astronomy Data And Computing Services (ADACS).
    
    Parts of this research were supported by the Australian Research Council Centre of Excellence for All Sky Astrophysics in 3~Dimensions (ASTRO~3D), through project number CE170100013.
    
    LC acknowledges support from the Australian Research Council Discovery Project and Future Fellowship funding schemes (DP210100337, FT180100066).
    
    PK acknowledges financial support by the German Federal Ministry of Education and Research (BMBF) under grant 05A17PC2 (Verbundprojekt D-MeerKAT-II).
    
    JMvdH acknowledges support from the European Research Council under the European Union's Seventh Framework Programme (FP/2007--2013) / ERC Grant Agreement nr.~291531 (HIStoryNU).
    
    This research has made use of the NASA/IPAC Extragalactic Database (NED), which is funded by the National Aeronautics and Space Administration and operated by the California Institute of Technology.
\end{acknowledgement}

\bibliography{wallaby_pilot}

\appendix

\section{Noise maps}
\label{sec_noise_maps}

\begin{figure*}[ht]
    \centering
    \includegraphics[width=0.8\linewidth]{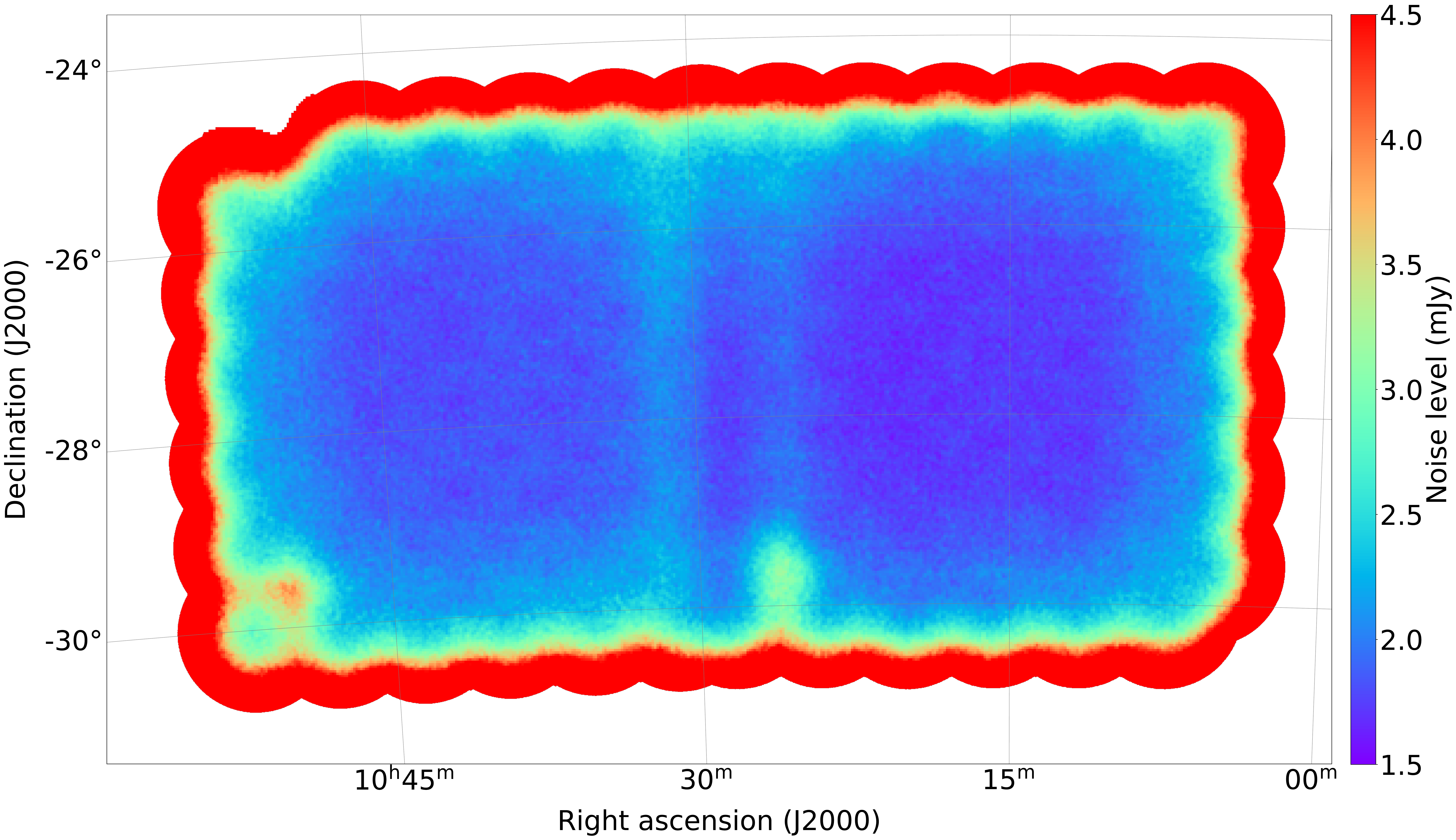}
    \caption{Noise variation across the Hydra field at a frequency of $\nu \approx 1418~\mathrm{MHz}$. Circular regions of increased noise indicate the locations of flagged primary beams, while the vertical structures seen near the centre of the field mark the overlap region between the two adjacent tiles making up the full field.}
    \label{fig_noise_hydra}
\end{figure*}

\begin{figure}[ht]
    \centering
    \includegraphics[width=\linewidth]{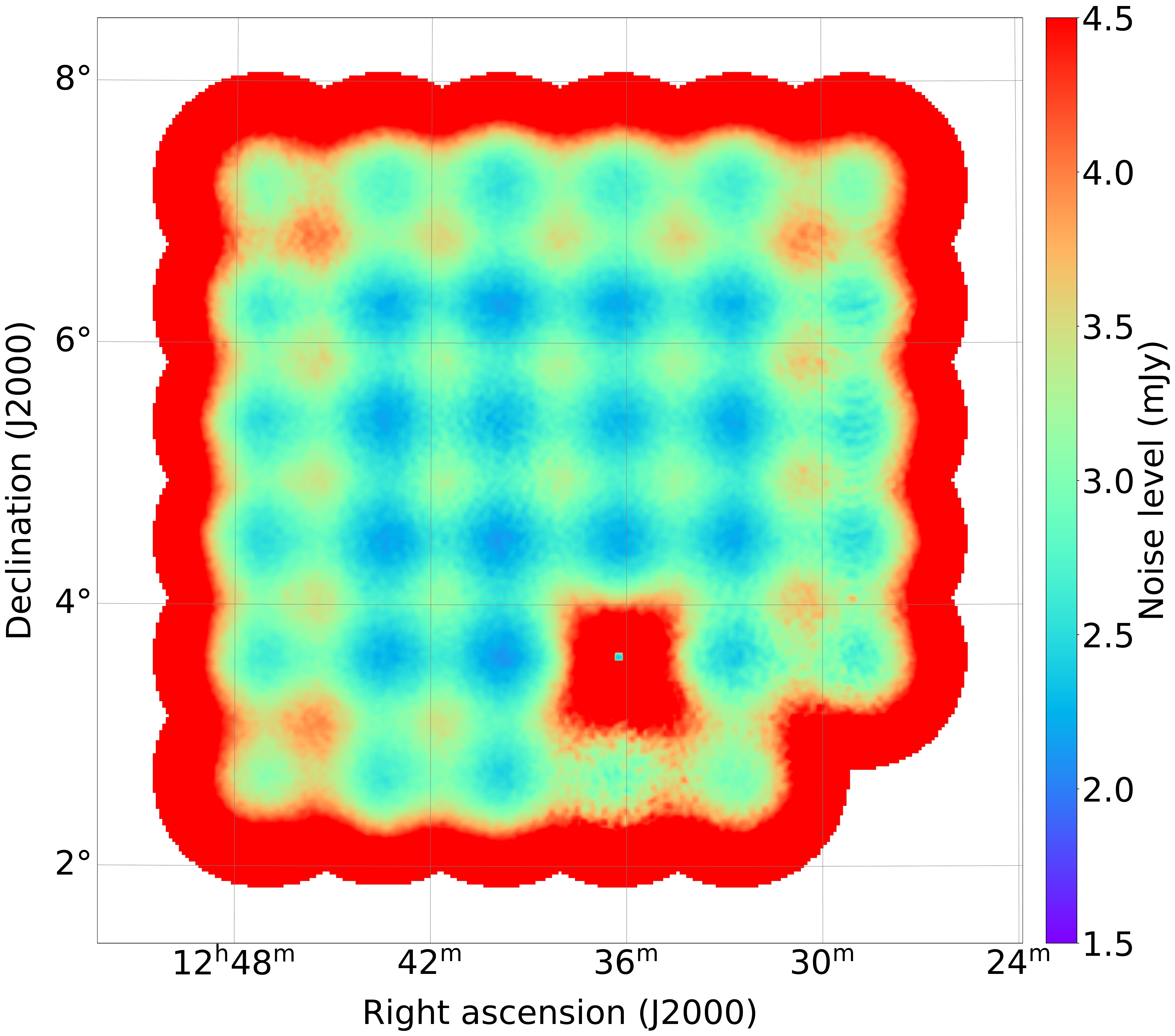}
    \includegraphics[width=\linewidth]{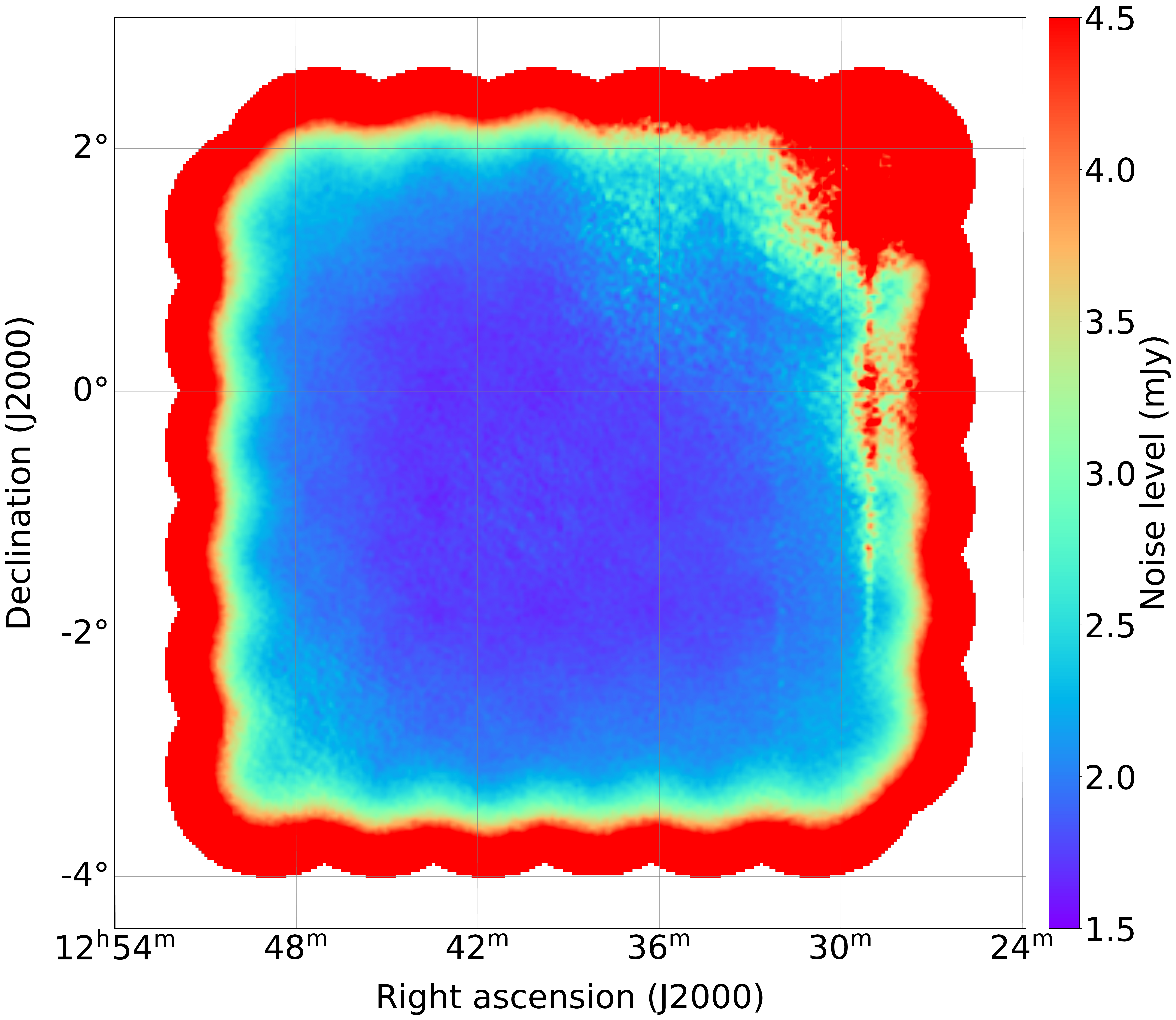}
    \caption{Noise variation across the northern (top) and southern (bottom) tile of the NGC~4636 field at a frequency of $\nu \approx 1418~\mathrm{MHz}$. Footprint~B of the northern tile was discarded due to poor data quality, while two additional primary beams from footprint~A were flagged by the data reduction pipeline. The increased noise level near the upper-right corner of the southern tile is due to residual continuum emission from the quasar 3C~273.}
    \label{fig_noise_n4636}
\end{figure}

\begin{figure}[ht]
    \centering
    \includegraphics[width=\linewidth]{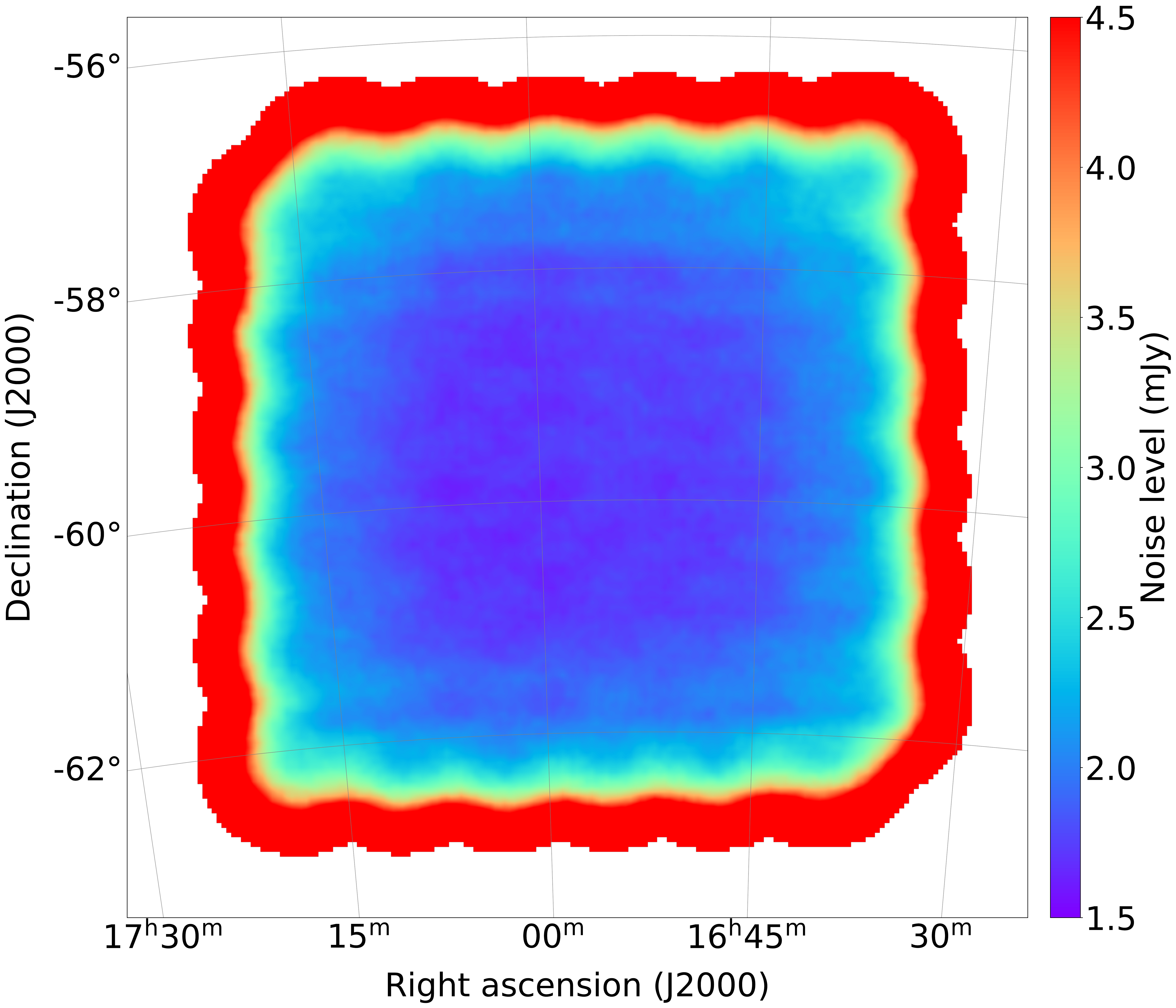}
    \caption{Noise variation across the eastern tile of the Norma field at a frequency of $\nu \approx 1418~\mathrm{MHz}$. The western tile of the Norma field is severely affected by radio continuum emission and not included in this data release.}
    \label{fig_noise_norma}
\end{figure}

Maps showing the variation of the RMS noise level across each field at a frequency of $\nu \approx 1418~\mathrm{MHz}$ are presented in Fig.~\ref{fig_noise_hydra}, \ref{fig_noise_n4636} and \ref{fig_noise_norma} for Hydra, NGC~4636 and Norma, respectively.

\section{Source catalogue and data products}
\label{app_data_products}

Table~\ref{tab_catalogue} lists all source parameters provided for each entry in the pilot survey catalogue, while Table~\ref{tab_products} gives an overview of the additional image products released for each individual entry in the catalogue.

\begin{table*}[t]
    \centering
    \caption{Source parameters included in the catalogue (see Section~\ref{sec_parameterisation} for further details). Note that the kinematic position angle measured by SoFiA is likely to differ from the position angle obtained from the kinematic modelling presented by \citet{Deg2022}.}
    \label{tab_catalogue}
    \begin{tabular}{lllll}
        \hline
        Parameter          & Type          & Unit    & Unified Content Descriptor (UCD)  & Description \\
        \hline
        \code{name} & \code{char} & -- & \code{meta.id;meta.main} & WALLABY source name (WALLABY Jhhmmss$\pm$ddmmss). \\
        \code{ra} & \code{double} & deg & \code{pos.eq.ra;meta.main} & Right ascension (J2000) of centroid position. \\
        \code{dec} & \code{double} & deg & \code{pos.eq.dec;meta.main} & Declination (J2000) of centroid position. \\
        \code{freq} & \code{double} & Hz & \code{em.freq;meta.main} & Barycentric frequency of centroid position. \\
        \code{f\_sum} & \code{double} & Jy\,Hz & \code{phot.flux;meta.main} & Integrated flux within 3D source mask. \\
        \code{err\_f\_sum} & \code{double} & Jy\,Hz & \code{stat.error;phot.flux} & Statistical uncertainty of integrated flux. \\
        \code{rms} & \code{double} & Jy/beam & \code{instr.det.noise} & Local RMS noise level near source. \\
        \code{w20} & \code{double} & Hz & \code{spect.line.width;meta.main} & Spectral line width at 20\% of the peak ($w_{20}$). \\
        \code{w50} & \code{double} & Hz & \code{spect.line.width} & Spectral line width at 50\% of the peak ($w_{50}$). \\
        \code{kin\_pa} & \code{double} & deg & \code{pos.posAng} & Position angle of kinematic major axis. \\
        \code{rel} & \code{double} & -- & \code{stat.probability} & Statistical reliability of detection from 0 to 1. \\
        \code{qflag} & \code{double} & -- & \code{meta.code.qual} & Source finding quality flag. \\
        \code{kflag} & \code{long} & -- & \code{meta.code.kin} & Kinematic model flag. \\
        \code{n\_pix} & \code{double} & -- & \code{meta.number;instr.pixel} & Number of pixels in 3D source mask. \\
        \code{f\_min} & \code{double} & Jy/beam & \code{phot.flux.density;stat.min} & Lowest flux density value within 3D source mask. \\
        \code{f\_max} & \code{double} & Jy/beam & \code{phot.flux.density;stat.max} & Highest flux density value within 3D source mask. \\
        \code{ell\_maj} & \code{double} & pix & \code{phys.angSize} & Major axis size of ellipse fitted to moment 0 map. \\
        \code{ell\_min} & \code{double} & pix & \code{phys.angSize} & Minor axis size of ellipse fitted to moment 0 map. \\
        \code{ell\_pa} & \code{double} & deg & \code{pos.posAng} & Position angle of ellipse fitted to moment 0 map. \\
        \code{ell3s\_maj} & \code{double} & pix & \code{phys.angSize} & Same as \code{ell\_maj}, but $> 3 \sigma$ pixels only and equal weights. \\
        \code{ell3s\_min} & \code{double} & pix & \code{phys.angSize} & Same as \code{ell\_min}, but $> 3 \sigma$ pixels only and equal weights. \\
        \code{ell3s\_pa} & \code{double} & deg & \code{pos.posAng} & Same as \code{ell\_pa}, but $> 3 \sigma$ pixels only and equal weights. \\
        \code{x} & \code{double} & pix & \code{pos.cartesian.x} & Centroid position in $x$.$^{\dagger}$ \\
        \code{err\_x} & \code{double} & pix & \code{stat.error;pos.cartesian.x} & Statistical uncertainty of centroid position in $x$. \\
        \code{y} & \code{double} & pix & \code{pos.cartesian.y} & Centroid position in $y$.$^{\dagger}$ \\
        \code{err\_y} & \code{double} & pix & \code{stat.error;pos.cartesian.y} & Statistical uncertainty of centroid position in $y$. \\
        \code{z} & \code{double} & pix & \code{pos.cartesian.z} & Centroid position in $z$.$^{\dagger}$ \\
        \code{err\_z} & \code{double} & pix & \code{stat.error;pos.cartesian.z} & Statistical uncertainty of centroid position in $z$. \\
        \code{x\_min} & \code{double} & pix & \code{pos.cartesian.x;stat.min} & Lower end of bounding box in $x$.$^{\dagger}$ \\
        \code{x\_max} & \code{double} & pix & \code{pos.cartesian.x;stat.max} & Upper end of bounding box in $x$.$^{\dagger}$ \\
        \code{y\_min} & \code{double} & pix & \code{pos.cartesian.y;stat.min} & Lower end of bounding box in $y$.$^{\dagger}$ \\
        \code{y\_max} & \code{double} & pix & \code{pos.cartesian.y;stat.max} & Upper end of bounding box in $y$.$^{\dagger}$ \\
        \code{z\_min} & \code{double} & pix & \code{pos.cartesian.z;stat.min} & Lower end of bounding box in $z$.$^{\dagger}$ \\
        \code{z\_max} & \code{double} & pix & \code{pos.cartesian.z;stat.max} & Upper end of bounding box in $z$.$^{\dagger}$ \\
        \code{dist\_h} & \code{double} & Mpc & \code{pos.distance} & Local, barycentric Hubble distance, $\mathrm{c}z/H_{0}$, for $H_{0} = 70~\mathrm{km \, s^{-1} \, Mpc^{-1}}$. \\
        \code{log\_m\_hi} & \code{double} & $M_{\sun}$ & \code{phys.mass} & \ion{H}{I} mass, $\log_{10}(M_{\ion{H}{I}} / M_{\sun})$, derived from local Hubble distance. \\
        \code{comments} & \code{char} & -- & \code{meta.note} & Comments on individual sources. \\
        \code{team\_release} & \code{char} & -- & \code{meta.dataset;meta.main} & Internal WALLABY team release identifier. \\

        \hline
        \multicolumn{5}{l}{$^{\dagger}\,$Pixel coordinates are relative to the region of the data cube processed by SoFiA and thus arbitrary.}
    \end{tabular}
\end{table*}

\begin{table*}[t]
    \centering
    \caption{Additional \ion{H}{I} data products supplied for each individual source from the catalogue.}
    \label{tab_products}
    \begin{tabular}{lllll}
        \hline
        Product  & File suffix        & Format     & Unit  & Description \\
        \hline
        Data     & \code{\_cube.fits} & FITS cube  & Jy/beam    & Small sub-cube around each source with extra padding of 10~pixels/channels. \\
        Mask     & \code{\_mask.fits} & FITS cube  & --    & Mask cube where source pixels are set to 1. \\
        Moment~0 & \code{\_mom0.fits} & FITS image & Jy/beam Hz & Integrated flux map from $0^{\rm th}$ spectral moment within mask. \\
        Moment~1 & \code{\_mom1.fits} & FITS image & Hz    & Line centroid map from $1^{\rm st}$ spectral moment of all positive pixels within mask. \\
        Moment~2 & \code{\_mom2.fits} & FITS image & Hz    & Line width map from $2^{\rm nd}$ spectral moment of all positive pixels within mask. \\
        Channels & \code{\_chan.fits} & FITS image & --    & Map of the number of spectral channels per pixel within mask. \\
        Spectrum & \code{\_spec.txt}  & Text file  & Jy    & Spectrum integrated across mask. \\
        \hline
    \end{tabular}
\end{table*}

\section{Comments on individual detections}
\label{sec_comments}

Table~\ref{tab_comments} lists comments on individual detections that may have been affected by issues such as continuum residuals, partial detection of a single galaxy or compound detections of multiple galaxies.

\begin{table*}[t]
    \centering
    \caption{Comments on individual detections ordered by field.}
    \label{tab_comments}
    \begin{tabular}{lll}
        \hline
        Field & Source & Comment \\
        \hline
        Hydra~TR1 & WALLABY J103758$-$252035 & Potential sidelobe. \\
                  & WALLABY J103732$-$261917 & Partial detection of galaxy; other half is WALLABY J103726$-$261843. \\
                  & WALLABY J103726$-$261843 & Partial detection of galaxy; other half is WALLABY J103732$-$261917. \\
                  & WALLABY J103442$-$283406 & Interacting system. \\
                  & WALLABY J103407$-$270622 & Partial detection of galaxy; other half is WALLABY J103405$-$270612. \\
                  & WALLABY J103405$-$270612 & Partial detection of galaxy; other half is WALLABY J103407$-$270622. \\
        \hline
        Hydra~TR2 & WALLABY J100321$-$291708 & Questionable. \\
                  & WALLABY J101049$-$302538 & Missing flux due to flagged continuum source. \\
                  & WALLABY J101434$-$274133 & Missing flux due to flagged continuum source. \\
                  & WALLABY J101443$-$263328 & Close galaxy pair? \\
                  & WALLABY J101448$-$274240 & Close galaxy pair? \\
                  & WALLABY J101934$-$261721 & Questionable. \\
                  & WALLABY J101945$-$272719 & Close galaxy pair? \\
                  & WALLABY J102019$-$285220 & Close galaxy pair? \\
                  & WALLABY J102054$-$263844 & Close galaxy pair? \\
                  & WALLABY J102207$-$282201 & Partial detection of galaxy? \\
                  & WALLABY J102447$-$264054 & Partial detection of galaxy; other half is WALLABY J102448$-$264152. \\
                  & WALLABY J102448$-$264152 & Partial detection of galaxy; other half is WALLABY J102447$-$264054. \\
                  & WALLABY J102605$-$280710 & Partial detection of galaxy; other half is WALLABY J102608$-$280840. \\
                  & WALLABY J102608$-$280840 & Partial detection of galaxy; other half is WALLABY J102605$-$280710. \\
                  & WALLABY J103405$-$270612 & Partial detection of galaxy; other half is WALLABY J103407$-$270622 from Hydra~TR1. \\
                  & WALLABY J103442$-$283406 & Interacting system of mutiple galaxies. \\
                  & WALLABY J103508$-$283427 & Questionable. \\
                  & WALLABY J103540$-$284607 & Close galaxy pair; WALLABY J103546$-$284602 and WALLABY J103539$-$284606 in Hydra~TR1. \\
                  & WALLABY J103543$-$255954 & Questionable. \\
                  & WALLABY J103726$-$261843 & Partial detection of galaxy; other half is WALLABY J103732$-$261920. \\
                  & WALLABY J103732$-$261917 & Partial detection of galaxy; other half is WALLABY J103725$-$261841. \\
                  & WALLABY J103809$-$260453 & Partial detection of galaxy? \\
                  & WALLABY J103818$-$285023 & Questionable. \\
                  & WALLABY J103853$-$274100 & Questionable. \\
                  & WALLABY J103858$-$300500 & Close galaxy pair? \\
        \hline
        NGC~4636~TR1 & WALLABY J122713$-$020236 & Partial detection of galaxy? \\
                     & WALLABY J122745+013601 & Affected by strong continuum residuals. \\
                     & WALLABY J123138+035620 & Affected by strong continuum residuals. \\
                     & WALLABY J123407+023905 & Affected by strong continuum residuals. \\
                     & WALLABY J123424+062511 & Interacting system and tidal streams. \\
                     & WALLABY J123427+021108 & Affected by strong continuum residuals and high noise near tile edge. \\
                     & WALLABY J123541$-$001253 & Interacting system? \\
                     & WALLABY J123734+042150 & Affected by flagged spectral channel range at low-frequency end. \\
                     & WALLABY J123905$-$022950 & Close galaxy pair? \\
                     & WALLABY J124109+041725 & Affected by strong continuum residuals. \\
                     & WALLABY J124318+015754 & \ion{H}{I} emission covers only small region of optical disc. \\
                     & WALLABY J124421+042536 & Close galaxy pair. \\
        \hline
        Norma~TR1 & WALLABY J164821$-$583425 & Questionable. \\
                  & WALLABY J165145$-$590915 & Part of gas bridge between NGC~6215 and NGC~6221. \\
                  & WALLABY J170550$-$620939 & Close galaxy pair. \\
                  & WALLABY J170753$-$602717 & Close galaxy pair. \\
                  & WALLABY J171828$-$574719 & Debris associated with WALLABY J171804$-$575135? \\
        \hline
    \end{tabular}
\end{table*}

\end{document}